**Kapitza thermal resistance in linear and nonlinear chain models: isotopic defect.**

O.V.Gendelman*, Jithu Paul

Faculty of Mechanical Engineering, Technion – Israel Institute of Technology,

Haifa, 3200003, Israel

* - contacting author, ovgend@technion.ac.il

**Abstract**

Kapitza resistance in the chain models with internal defects is considered. For the case of the linear chain, the exact analytic solution for the boundary resistance is derived for arbitrary linear time-independent conservative inclusion or defect. A simple case of isolated isotopic defects is explored in more detail. Contrary to the bulk conductivity in the linear chain, the Kapitza resistance is finite. However, the universal thermodynamic limit does not exist in this case. In other terms, the exact value of the resistance is not uniquely defined, and depends on the way of approaching the infinite lengths of the chain fragments. By this reason, and also due to the explicit dependence on the parameters of the thermostats, the resistance cannot be considered as a local property of the defect. Asymptotic scaling behavior of the heat flux in the case of very heavy defect is explored and compared to the nonlinear counterparts; similarities in the scaling behavior are revealed. For the lightweight isotopic defect in the linear chain, one encounters a typical dip of the temperature profile, related to weak excitation of the localized mode in the attenuation zone. If the nonlinear interactions are included, this dip can still appear at a relatively short time scale, with subsequent elimination due to the nonlinear interactions. This observation implies that even in the nonlinear chains, the linear dynamics can predict the main features of the short-time evolution of the thermal profile if the temperature is low enough.

## 1. Introduction

The history of thermal boundary resistance starts from the famous work of Kapitza on temperature discontinuity at the interface between a metal and liquid Helium II [1]. Multiple and versatile experimental results on the boundary thermal resistance are available in the literature [2–5]. The first attempt to offer a theoretical explanation based on the phonon transmission coefficient, particularly at low-temperature experiments, was made by Khalatnikov in 1952 [6] and by Little in 1959 [7], and it is generally known as the acoustic mismatch model (AMM). The well-known limitations of AMM are that, first; it fails to predict the Kapitza resistance when the contacting structures are similar. Then, it considers only long-wavelength phonons since the theory has been developed for low-temperature experiments. The first drawback was somewhat corrected by including a multicomponent phonon distribution functions on each side of the interface [8]. Additional developments included account of the phonon scattering at and near the interface [9]. Several other works were based on transport theory approach, but complexity of the calculations limited its practical use [10–12]. Sluckin developed a simple model for the chain of billiard particles and the results were surprisingly matching with experiments [13]. There were attempts to extend Khalatnikov's theory to harmonic systems [14,15]; in Ref. [15] the results were compared with molecular dynamics simulations. In 1987, Swartz and Pohl [16] proposed a diffuse mismatch model to explain the higher temperature behavior of thermal boundary resistance where the AMM becomes ineffective. This theory considers other extremes compared to the AMM, that is, all the phonons are considered as diffusely scattered at the interface, and any acoustic correlation at the interface is destroyed.

Harmonic crystals are the simplest model to study thermal properties of dielectrics. It is well-known that the dynamics of homogeneous harmonic chain in modal coordinates reduces to an assembly of freely moving phonons, therefore the thermal conductivity in this model is anomalous. The modern era studies of the heat transport in harmonic chain started from the famous work of Rieder-Lebowitz-Lieb (RLL) [17] that presented the exact solution for one-dimensional homogeneous chain coupled to Langevin thermostats. It was demonstrated that the heat flux is proportional to the temperature difference between the boundaries. Besides, the exponential boundary layer near the thermostats has been observed. Then the studies progressed into more complicated problems that involved various scattering mechanisms by inserting various impurities like defect mass or defective coupling, or the phonon-phonon scattering caused by the nonlinearity [18]. In Ref. [19], the problem of infinite disordered harmonic chain has been analyzed. It was demonstrated that the heat flux is proportional to the boundary temperature difference, provided that the spectral measure exhibits an absolutely continuous part. For the special case of alternating-mass-chain, the temperature profile oscillates through the chain as follows from the exact solution in Ref. [20]. Somewhat surprisingly, such temperature oscillations also revealed themselves in the alternate-mass Fermi-Pasta-Ulam (FPU) model [21].

Thermal transport in the linear chain with isotopic defect has been extensively studied numerically in Ref. [22]. A recent paper [23]- explored numerically the Kapitza resistance in a variety of benchmark models with isolated isotopic defect. Current work revisits this problem, and presents exact analytic solution for the Kapitza resistance in the linear chain with isolated defect (interface) in the limit of infinite chain length.

This exact solution allows comprehensive study of the boundary resistance in the linear chain, including the near-field around the defect. Besides, the heat flux exhibits interesting asymptotic behavior in the limit of large defect mass. This behavior is compared to numeric data concerning similar asymptotic limit for *β-FPU* and Lennard-Jones (LJ) chains.

If the mass of the inclusion is less than the mass of other chain particles, the thermal profile exhibits an interesting "dip", indicating a "cold" point in the chain. Such behavior is paradoxical, at odds with the Second Law of thermodynamics – the heat flows from the "cold" point defect to a hotter part of the chain. Such anomalous behavior is a well-known artifact of the linear model; similar behavior is observed also at larger scales [24,25]. It is demonstrated below that account of the nonlinearity removes this anomaly. More exactly, when the temperature gradient is imposed, the dip is initially formed also in the nonlinear chains. Then the dip is gradually destroyed, ending up with monotonous thermal profile. Possible role of mobile breathers [26,27] in destruction of the dip is discussed.

The general outline of this paper is as follows. Section 2 presents the exact solution for the Kapitza resistance in the linear chain containing linear time-independent (LTI) inclusion. In Section 3, the expressions for special case-of the isotopic defect are derived and asymptotic limit for the heat flux in the case of very heavy defect is compared to similar phenomena in $\beta - FPU$ and LJ models. Section 4 addresses in detail the thermal "dip" at light defect in the linear model and its connection to localized mode in the chain spectrum. The treatment is extended to the $\beta - FPU$ chain, and disappearance of the dip is explored numerically. Then, the thermal near-field of the defect in linear model is explored and exponential convergence to the average temperature is established. The exponent is related to the defect mass. Section 5 is devoted to discussion and concluding remarks.

## 2. Kapitza resistance in linear chain with inclusion: general treatment.

Let us consider a heat transport in a linear chain with linear time-independent (LTI) inclusion, with thermostats attached to the terminal particles at the right and at the left. A general sketch of the model system is presented in Fig. 1.

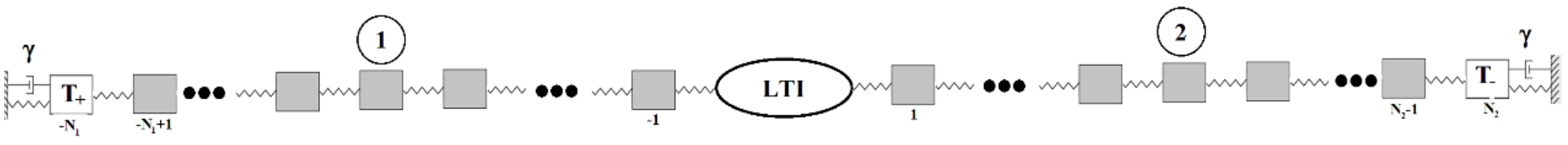

FIG. 1. Sketch of the model system

All masses and stiffnesses within the chain fragments beyond the inclusion are set to unity without affecting the generality. The length of the fragments 1 and 2 is set to $N_{1,2}$ respectively. At this stage, the structure of the inclusion, or the defect, in the chain is left generic. The only assumption is that the motion of all particles that belong to the inclusion is described by linear time-independent equations. Beyond the inclusion and apart from the terminal particles, the chain is described by common equations of motion:

$$\ddot{u}_n + 2u_n - u_{n-1} - u_{n+1} = 0 \tag{1}$$

Traveling waves in this chain obey a well-known dispersion relation:

$$\omega = 2\sin\left(q/2\right), \quad -\pi \le q \le \pi \tag{2}$$

Here $\omega(q)$ denotes the frequency of the traveling wave, *q* is the wave vector. The system is excited through thermostats attached to the terminal particles. For this sake, we assume that both these particles are subject to linear viscous damping with coefficient $\gamma$ and excited by stochastic forces. Appropriate equations of motion are written as follows:

$$\begin{aligned} &\ddot{u}_{-N_1} + 2u_{-N_1} - u_{-N_1+1} + \gamma\dot{u}_{-N_1} = \xi_+(t) \\ &\ddot{u}_{N_2} + 2u_{N_2} - u_{N_2-1} + \gamma\dot{u}_{N_2} = \xi_-(t) \end{aligned} \tag{3}$$

At this stage, the forcing functions in (3) will also stay generic, under assumption of zero mean and lack of correlation:

$$\langle \xi_\pm(t) \rangle = 0, \ \langle \xi_+(t_1)\xi_-(t_2) \rangle = 0 \tag{4}$$

Moreover, to simplify the expressions, we set $\xi_+(t) = \xi(t), \ \xi_-(t) = 0$. The care of original problem may be taken by appropriate superposition.

For sufficiently long chains, one can admit that the heat flux is realized only by waves that belong to the propagation zone of the chain. Thus, for fragments of the chain far from the terminal particles, the displacements of the particles can be expressed as linear combinations of the waves travelling to the left and to the right, with the frequencies belonging to the propagation zone:

$$\begin{aligned} &u_n = \int_P \exp(i\omega t)(\alpha_1(\omega)\exp(-iqn) + \beta_1(\omega)\exp(iqn))d\omega \quad \text{(Fragment 1)} \\ &u_n = \int_P \exp(i\omega t)(\alpha_2(\omega)\exp(-iqn) + \beta_2(\omega)\exp(iqn))d\omega \quad \text{(Fragment 2)} \end{aligned} \tag{5}$$

In the following, $\omega$ - and $q$ - dependence of all functions will be suppressed, where it will not cause the misunderstanding. The temperature is defined as a double average kinetic energy of the particle. If one assumes absence of phase correlations between the waves, the energy density transported in the right and left directions in the fragments is evaluated as $\rho_L(\omega)=\frac{\omega^2}{2}\left|\alpha_i(\omega)\right|^2$ and $\rho_R(\omega)=\frac{\omega^2}{2}\left|\beta_i(\omega)\right|^2$ respectively, $i=1,2$. The energy is transported with group velocity $v_{gr}(\omega)$. Then, with account of dispersion relation (2) and symmetry $\omega\to-\omega$ ,the temperatures of the chain fragments far enough from the thermostats and the inclusion, and heat fluxes through the fragments ($T_1$, $T_2$, $J_1$ and $J_2$ respectively) are expressed as follows:

$$T_1=\int_0^2\omega^2\left(\left|\alpha_1\right|^2+\left|\beta_1\right|^2\right)d\omega,\ J_1=\int_0^2\omega^2\left(\left|\alpha_1\right|^2-\left|\beta_1\right|^2\right)\left|v_{gr}\right|d\omega;$$
$$T_2=\int_0^2\omega^2\left(\left|\alpha_2\right|^2+\left|\beta_2\right|^2\right)d\omega,\ J_2=\int_0^2\omega^2\left(\left|\alpha_2\right|^2-\left|\beta_2\right|^2\right)\left|v_{gr}\right|d\omega; \qquad (6)$$
$$\left|v_{gr}\right|=\left|\frac{d\omega}{dq}\right|=\left|\cos\frac{q}{2}\right|$$

Note that the inclusion can, in principle, include the damping elements, and therefore the heat fluxes in fragments 1 and 2 are not necessarily equal.

Equations (3) lead to the following boundary conditions for Fourier components of the heat flux in the propagation zone:

$$\begin{aligned}&\alpha_1\exp(iqN_1)c_+ +\beta_1\exp(-iqN_1)c_- =\Xi\\&\alpha_2\exp(-iqN_2)c_- +\beta_2\exp(iqN_2)c_+ =0\\&c_\pm=\exp(\pm iq)+i\gamma\omega\end{aligned} \qquad (7)$$

Here $\Xi$ denotes the Fourier transform of the forcing function $\xi(t)$. To close System (7) we recall that the inclusion is assumed linear and time-independent. Therefore, the amplitudes of incoming and outgoing waves in the chain fragments are related through certain transfer-matrix, defined as

$$\begin{pmatrix}\alpha_1\\ \beta_1\end{pmatrix}=\mathbf{G}\begin{pmatrix}\alpha_2\\ \beta_2\end{pmatrix},\ \mathbf{G}=\begin{pmatrix}g_{11} & g_{12}\\ g_{21} & g_{22}\end{pmatrix} \qquad (8)$$

Then, system (7) is rewritten in the form

$$\begin{aligned}&\alpha_2(g_{11}\exp(iqN_1)c_+ +g_{21}\exp(-iqN_1)c_-)+\beta_2(g_{12}\exp(iqN_1)c_+ +g_{22}\exp(-iqN_1)c_-)=\Xi\\&\alpha_2\exp(-iqN_2)c_- +\beta_2\exp(iqN_2)c_+ =0\end{aligned} \qquad (9)$$

System (9) is solved straightforwardly:

$$\alpha_2=\frac{\Xi\exp(iqN_2)c_+}{D},\beta_2=-\frac{\Xi\exp(-iqN_2)c_-}{D}$$
$$D=g_{11}\exp(iq(N_1+N_2))c_+^2+(g_{21}\exp(iq(N_2-N_1))-g_{12}\exp(iq(N_1-N_2)))c_+c_- - \quad (10)$$
$$-g_{22}\exp(-iq(N_1+N_2))c_-^2$$

To achieve certain simplification, it is instructive to explore the properties of transfer matrix $\mathbf{G}$ in more depth. Let us consider the reciprocal system, in which the forcing is applied to the right end of the chain. Then, System (9) is substituted by the following equations:

$$\tilde{\alpha}_2(g_{11}\exp(iqN_1)c_+ + g_{21}\exp(-iqN_1)c_-)+\tilde{\beta}_2(g_{12}\exp(iqN_1)c_+ + g_{22}\exp(-iqN_1)c_-)=0$$
$$\tilde{\alpha}_2\exp(-iqN_2)c_- + \tilde{\beta}_2\exp(iqN_2)c_+ = \Xi$$

(11)

According to Raleigh reciprocity theorem [28], the displacement of the rightmost particle according to System (9) is equal to the displacement of the leftmost particle according to system (11):

$$\alpha_2\exp(-iqN_2)+\beta_2\exp(iqN_2)=\tilde{\alpha}_1\exp(iqN_1)+\tilde{\beta}_1\exp(-iqN_1);\ \begin{pmatrix}\tilde{\alpha}_1\\ \tilde{\beta}_1\end{pmatrix}=\mathbf{G}\begin{pmatrix}\tilde{\alpha}_2\\ \tilde{\beta}_2\end{pmatrix} \quad (12)$$

From (10-12), after some simple algebra, one obtains:

$$g_{11}g_{22}-g_{12}g_{21}=\det\mathbf{G}=1 \quad (13)$$

This conclusion is valid, even if the inclusion is not symmetric and contains the damping elements, i.e. is not conservative. Further simplification is achieved if one assumes that the inclusion is conservative. In this case, due to the energy conservation, in the stationary regime the heat fluxes through fragments 1 and 2 are equal. Then, for any forcing configuration the relationship

$$|\alpha_1|^2-|\beta_1|^2=|\alpha_2|^2-|\beta_2|^2 \quad (14)$$

must hold identically. Therefore, the elements of the transfer matrix obey the following relationships:

$$|g_{11}|^2-|g_{21}|^2=1;\ |g_{22}|^2-|g_{12}|^2=1;\ g_{11}g_{12}^*=g_{21}g_{22}^* \quad (15)$$

It is easy to derive that the most generic transfer matrix that respects conditions (13) and (15) may be represented with the help of the following convenient parametrization:

$$\mathbf{G}=\begin{pmatrix}\cosh x\exp(i\theta_1) & \sinh x\exp(-i\theta_2)\\ \sinh x\exp(i\theta_2) & \cosh x\exp(-i\theta_1)\end{pmatrix};\ x\in[0,\infty),\theta_1,\theta_2\in[-\pi,\pi) \quad (16)$$

For further derivations, we will assume that the inclusion does not contain the damping elements, and therefore the transfer matrix is described by parametrization (16). It is interesting to note that from Eq. (15) one can derive that $\left|\det \mathbf{G}\right| = 1$, but cannot obtain stronger condition (13). With account of solution (10) and parametrization (16), expressions (6) for the temperatures and the heat flux in the case of conservative inclusion are presented as follows:

$$
\begin{aligned}
&J = \int_P \omega^2 \left|\Xi\right|^2 \left|v_{gr}\right| \frac{\left|c_+\right|^2 - \left|c_-\right|^2}{\left|D\right|^2} d\omega; \\
&T_1 = \int_P \frac{\omega^2 \left|\Xi\right|^2}{\left|D\right|^2} \left[ \begin{array}{l} (\cosh^2 x + \sinh^2 x)(\left|c_+\right|^2 + \left|c_-\right|^2) - \sinh 2x \times \\ \times (c_+ c_-^* \exp i(\theta_1 + \theta_2 + 2qN_2) + c_+^* c_- \exp i(-\theta_1 - \theta_2 - 2qN_2)) \end{array} \right] d\omega; \\
&T_2 = \int_P \omega^2 \left|\Xi\right|^2 \frac{\left|c_+\right|^2 + \left|c_-\right|^2}{\left|D\right|^2} d\omega; \\
&D = \cosh x \exp i(q(N_1 + N_2) + \theta_1) c_+^2 + \sinh x c_+ c_- (\exp i(q(N_2 - N_1) + \theta_2) - \\
&- \exp i(q(N_1 - N_2) - \theta_2)) - \cosh x \exp i(-\theta_1 - 2q(N_1 + N_2)) c_-^2
\end{aligned}
\tag{17}
$$

The limit of infinite system $N_{1,2} \to \infty$ involves averaging over rapidly oscillating phases $\varphi_{1,2} = qN_{1,2}$ in the exponents for fixed values of the wavenumber. However, the presence of two not necessarily equal chain fragments leads to a mathematical peculiarity absent in homogeneous systems [17–19]. The universal thermodynamic limit of the infinite system size *does not exist*. The reason is that in order to perform the averaging, one should specify the ratio $r = \lim\limits_{\substack{N_1 \to \infty \\ N_2 \to \infty}} (N_1 / N_2)$. If *r* is irrational, zero or infinite, the averaging over the two rapidly oscillating phases for all functions including these variables can be performed independently, and for arbitrary integrable function *F* is defined as

$$
\left\langle F(\exp(i\varphi_1), \exp(i\varphi_2)) \right\rangle_{\substack{N_1 \to \infty \\ N_2 \to \infty}} = \frac{1}{4\pi^2} \int_0^{2\pi} d\varphi_1 \int_0^{2\pi} d\varphi_2 F(\exp(i\varphi_1), \exp(i\varphi_2)) \tag{18}
$$

We refer to this case as *non-resonant*. For the considered system, details of the straightforward but somewhat awkward evaluation (17-18) are presented in Appendix A. The average heat flux and temperatures in the non-resonant case are expressed as follows:

$$J=\int_P \frac{\omega^2 |\Xi|^2 |v_{gr}| d\omega}{\cosh^2 x\sqrt{\left(|c_+|^2+|c_-|^2\right)^2-4\tanh^2 x|c_+|^2|c_-|^2}}$$

$$T_2=\int_P \frac{\omega^2 |\Xi|^2 \left(|c_+|^2+|c_-|^2\right) d\omega}{\cosh^2 x\left(|c_+|^2-|c_-|^2\right)\sqrt{\left(|c_+|^2+|c_-|^2\right)^2-4\tanh^2 x|c_+|^2|c_-|^2}} \qquad (19)$$

$$T_1=T-T_2$$

If the ratio *r* is rational, we denote $r=m/n$ , where *m,n* are integers without the common divisors. This case is referred to as $m:n$ *resonance*. The averaging over two rapidly oscillating phases is not independent, since these phases are locked:

$$\left\langle F(\exp(i\varphi_1),\exp(i\varphi_2))\right\rangle_{\substack{N_1\to\infty\\N_2\to\infty}}=\frac{1}{2\pi}\int_0^{2\pi} d\psi F(\exp(im\psi),\exp(in\psi)),\quad \psi=\varphi_2/n \qquad (20)$$

This latter case yields, generally, more complicated integrals. Still, for the simplest case of 1:1 resonance, the averaging (20) can be performed explicitly. The heat flux and the average temperatures at both sides of the inclusion are presented in the following general form (see Appendix A for the derivation details):

$$J=\int_P \omega^2|\Xi|^2|v_{gr}|\frac{|c_+|^2+|c_-|^2}{\cosh^2 x\left(|c_+|^4+|c_-|^4+2|c_+|^2|c_-|^2\left(1-2\tanh^2 x\sin^2\theta_2\right)\right)}d\omega;$$

$$T_1=\int_P \omega^2|\Xi|^2\frac{(\cosh^2 x+\sinh^2 x)(|c_+|^2+|c_-|^2)^2-8|c_+|^2|c_-|^2\sinh^2 x\sin^2\theta_2}{\cosh^2 x\left(|c_+|^2-|c_-|^2\right)\left(|c_+|^4+|c_-|^4+2|c_+|^2|c_-|^2\left(1-2\tanh^2 x\sin^2\theta_2\right)\right)}d\omega;$$

$$T_2=\int_P \omega^2|\Xi|^2\frac{\left(|c_+|^2+|c_-|^2\right)^2}{\cosh^2 x\left(|c_+|^2-|c_-|^2\right)\left(|c_+|^4+|c_-|^4+2|c_+|^2|c_-|^2\left(1-2\tanh^2 x\sin^2\theta_2\right)\right)}d\omega;$$

(21)

Below we refer to this case of 1:1 case simply as *resonant*, and specify other resonances if necessary.

Let us adopt, in addition, that the forcing function in Eq. (3) corresponds to white Gaussian noise, i.e. the self-correlations obey the well-known relations

$$\left\langle \xi_\pm(t_1)\xi_\pm(t_2)\right\rangle=2\gamma T_\pm\delta(t_1-t_2) \qquad (22)$$

Then one sets $T_+=T, T_-=0, |\Xi|^2=\frac{\gamma T}{\pi}$. For the trivial case of the homogeneous chain the transfer matrix is unit, therefore $x=0$, and from Eq. (19) and (21) one obtains the following well-known results [17]:

$$T_1 = T_2 = T/2$$
$$J_h = \frac{T}{4\gamma^3}\left(1 + 2\gamma^2 - \sqrt{1+4\gamma^2}\right) \tag{23}$$

As one can expect, for the system without defect there is no difference between the resonant and non-resonant cases. Further treatment requires one to specify the considered inclusion or defect. In the next Section we consider the case of a single isotopic defect in the chain.

## 3. Kapitza resistance in the case of a single isotopic defect.

### 3.1 Linear chain

Assume that the only inhomogeneity in the chain is the single isotopic defect – a particle with mass $m \neq 1$at the site $n = 0$. The chain (beyond the thermostats) is therefore described by the following equations:

$$(1 + \delta_{no}(m-1))\ddot{u}_n + 2u_n - u_{n-1} - u_{n+1} = 0 \tag{24}$$

Then, Fragment 1 corresponds to all particles with n<0, and fragment 2 – to all particles with n>0.

Substituting expansions (5) into (24), one obtains the following relationships:

$$\begin{aligned} &\alpha_1 + \beta_1 = \alpha_2 + \beta_2 \\ &\exp(iq)\alpha_1 + \exp(-iq)\beta_1 = (2 - m\omega^2 - \exp(-iq))\alpha_2 + (2 - m\omega^2 - \exp(iq))\beta_2 \end{aligned} \tag{25}$$

Therefore, the transfer matrix is expressed as follows:

$$\begin{aligned} \mathbf{G} &= \begin{pmatrix} 1 & 1 \\ e^{iq} & e^{-iq} \end{pmatrix}^{-1} \begin{pmatrix} 1 & 1 \\ 2 - m\omega^2 - e^{-iq} & 2 - m\omega^2 - e^{iq} \end{pmatrix} = \\ &= \frac{i}{2\sin q} \begin{pmatrix} (m-1)\omega^2 - 2i\sin q & (m-1)\omega^2 \\ -(m-1)\omega^2 & -2i\sin q - (m-1)\omega^2 \end{pmatrix} \end{aligned} \tag{26}$$

The values necessary for evaluation of integrals (19) and (21) are expressed as:

$$\sinh^2 x = \frac{(m-1)^2\omega^4}{4\sin^2 q} = \frac{(m-1)^2\omega^2}{4-\omega^2}, \theta_2 = -\frac{\pi}{2}\operatorname{sgn}\left(\frac{m-1}{\sin q}\right) \tag{27}$$

Substituting (27) into integrals (19) and denoting $z = \omega/2$, one obtains the following expressions for the non-resonant case:

$$J=\frac{8\gamma T}{\pi}\int_0^1\frac{z^2\left(1-z^2\right)dz}{\sqrt{1+z^2\left((m-1)^2-1\right)}\sqrt{\left(1+4\gamma^2z^2\right)^2+64(m-1)^2\gamma^2z^6}}$$

$$T_2=\frac{T}{\pi}\int_0^1\frac{\left(1+4\gamma^2z^2\right)dz}{\sqrt{1+z^2\left((m-1)^2-1\right)}\sqrt{\left(1+4\gamma^2z^2\right)^2+64(m-1)^2\gamma^2z^6}} \tag{28}$$

$$T_1=T-T_2$$

For the resonant case, one obtains from (21,27):

$$J=\frac{8\gamma T}{\pi}\int_0^1\frac{z^2\sqrt{1-z^2}(1+4\gamma^2z^2)dz}{(1+4\gamma^2z^2)^2+64(m-1)^2\gamma^2z^6}$$

$$T_1=\frac{T}{\pi}\int_0^1\frac{((1+4\gamma^2z^2)^2+128(m-1)^2\gamma^2z^6)dz}{\left((1+4\gamma^2z^2)^2+64(m-1)^2\gamma^2z^6\right)\sqrt{1-z^2}}=$$

$$=\frac{T}{2}+\frac{64(m-1)^2\gamma^2T}{\pi}\int_0^1\frac{z^6dz}{\left((1+4\gamma^2z^2)^2+64(m-1)^2\gamma^2z^6\right)\sqrt{1-z^2}} \tag{29}$$

$$T_2=\frac{T}{\pi}\int_0^1\frac{(1+4\gamma^2z^2)^2dz}{\left((1+4\gamma^2z^2)^2+64(m-1)^2\gamma^2z^6\right)\sqrt{1-z^2}}=$$

$$=\frac{T}{2}-\frac{64(m-1)^2\gamma^2T}{\pi}\int_0^1\frac{z^6dz}{\left((1+4\gamma^2z^2)^2+64(m-1)^2\gamma^2z^6\right)\sqrt{1-z^2}}$$

Integrals in (29) can be explicitly evaluated since the denominators contain a cubic polynomial with respect to $z^2$. The details of somewhat awkward calculation are given in Appendix B.

Now we illustrate numerically the effect of various resonances on the value of the Kapitza resistance. In all cases, the latter is evaluated as

$$R_K=\frac{T_1-T_2}{J}\ . \tag{30}$$

To this end, the heat flux in the linear chain with $N=1001$ particles and single isotopic defect at different sites is evaluated by numerical solution of the RLL problem [17,22]. The dependence of the resistance on the defect placement is depicted in Fig. 2.

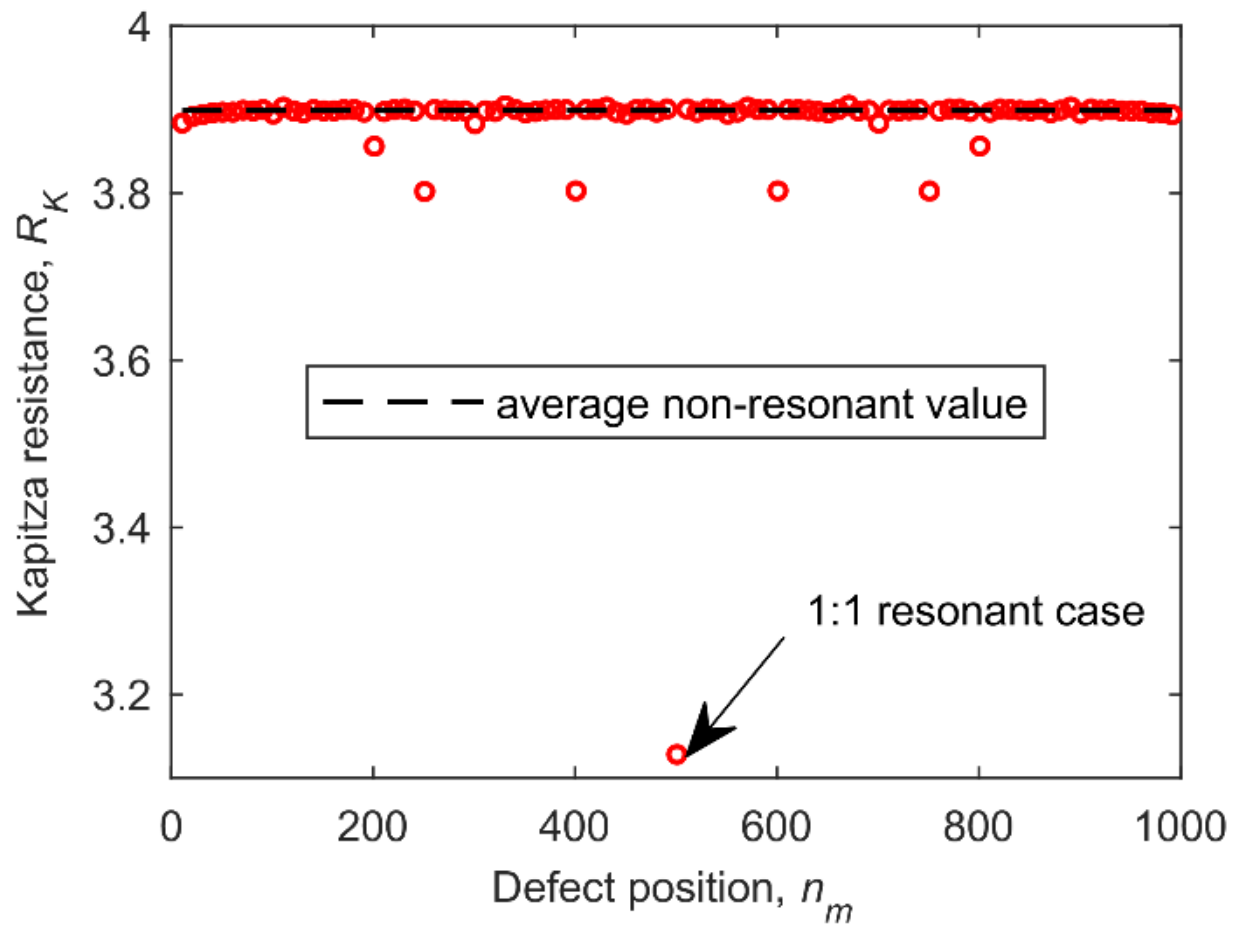

FIG. 2. (Color online) The Kapitza resistance for various positions of the isotopic defect in the chain with fixed length. Horizontal line corresponds to the non-resonant background value according to (28), $T = 1, m = 2, \gamma = 1, N = 1001$.

One can observe that for most placements of the defect the measured value of the Kapitza resistance is very close to the background non-resonant value defined by expressions (28). There are some outliers for "strong" resonances (1:3, 2:3), the strongest is the case of 1:1 resonance. Variation of the value of Kapitza resistance due to the resonances achieves about 20% in this particular case.

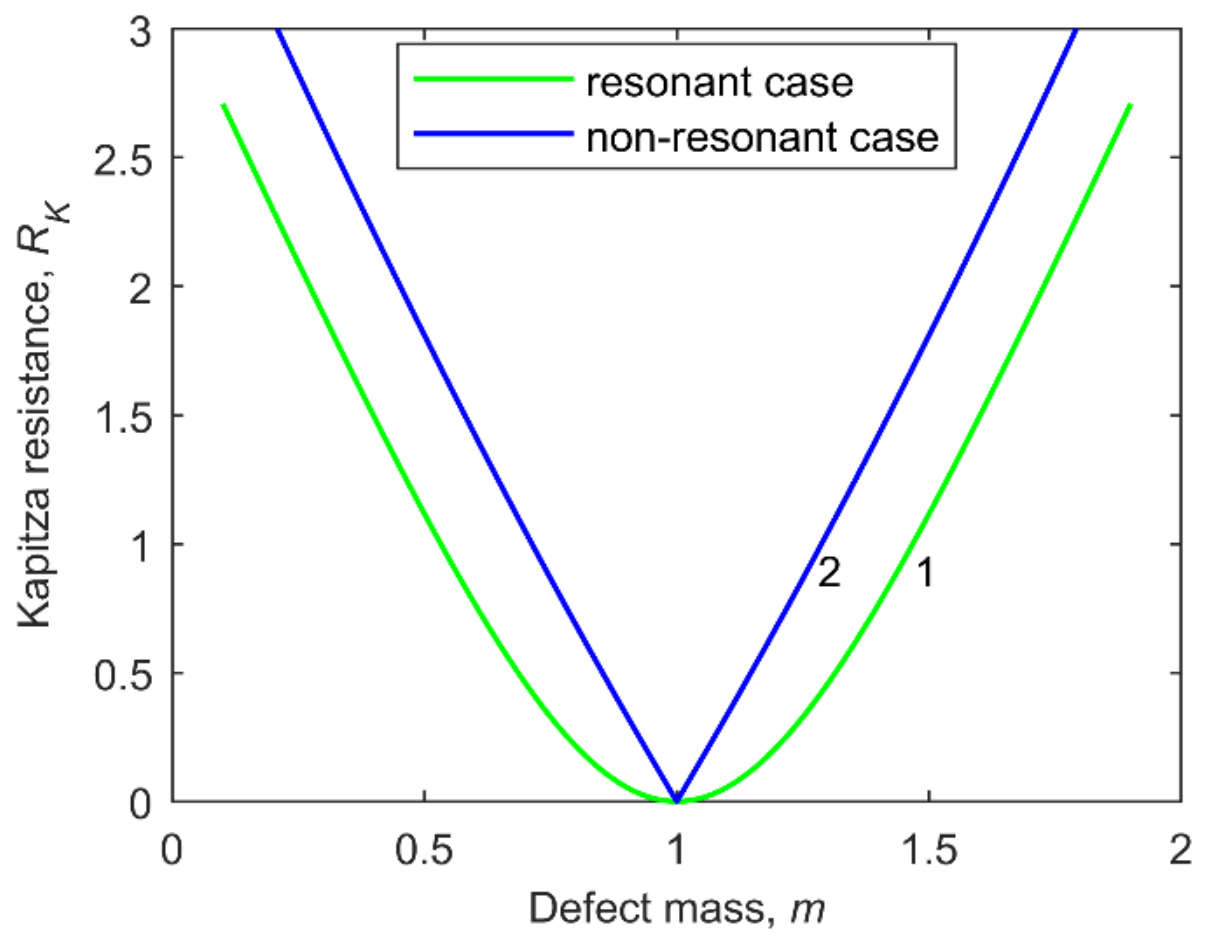

FIG. 3. (Color online) Analytical results of Kapitza resistance plotted by varying the defect mass. Marker 1 shows the resonant case and marker 2 shows the non-resonant case. $T = 1, \gamma = 1$

Dependency of the Kapitza resistance (30) on the defect mass is presented in Fig. 3, and on the coupling friction - in Fig. 4.

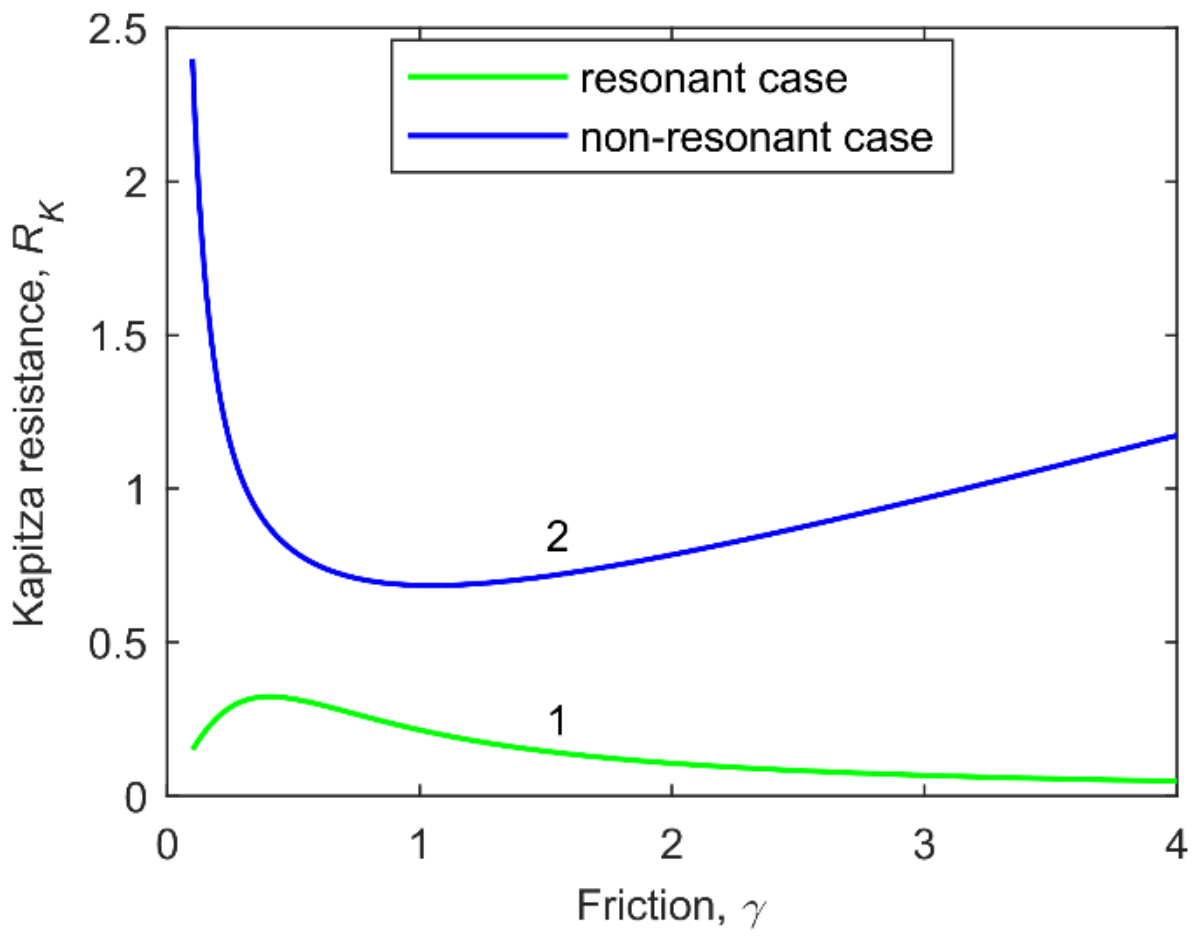

FIG. 4. (Color online) Analytical results of Kapitza resistance plotted by varying the coupling friction. Marker 1 shows the resonant case and marker 2 shows the non-resonant case. $T=1, m=1.2$

Fig. 3 indicates that, contrary to the bulk conductivity, the Kapitza resistance is finite even in the linear chain. From Figs 2-4, one learns that the resistance is not local property of the defect – it depends on particular way of taking the thermodynamic limit, and strongly, and even non-monotonously depends on the friction coefficient in the thermostats. Therefore, one requires nonlinear models to get in line with basic physical intuition concerning the locality of the boundary resistance [23].

### 3.2 Asymptotic behavior of the boundary resistance.

It is instructive to explore asymptotic behavior of the heat flux and the Kapitza resistance in the special cases of weak and heavy isotopic defects. The case of *weak defect* is defined as $|m-1| \ll 1, \gamma \ll 1$. In this limit, for the non-resonant case, expressions (28, 30) yield

$$T_1 - T_2 \approx \frac{2T}{\pi}|m-1|; J \approx \frac{1}{2}\gamma T; R_K \approx \frac{4|m-1|}{\pi\gamma} \tag{31}$$

Therefore, the dependence is *non-analytic* as $|m-1| \to 0$, similarly to the behavior observed in Fig. 3 . For the resonant case, expressions (29, 30) yield

$$T_1 - T_2 \approx 20\gamma^2 (m-1)^2 T\,;\ J \approx \tfrac{1}{2}\gamma T\,; R_K \approx 40\gamma(m-1)^2 \tag{32}$$

One observes that in the resonant case the resistance *analytically* depends on the mass mismatch (cf. Fig. 3). Thus, the resonant and non-resonant cases exhibit qualitatively different asymptotic behavior.

The case of *heavy isotopic defect* is defined as $\gamma \ll 1, |m-1|\gamma \gg 1$. For the non-resonant case, denoting $\hat{z} = mz$, one obtains the following estimation:

$$J \approx \frac{8\gamma T}{\pi m^3} \int_0^m \frac{\hat{z}^2}{\sqrt{1+\hat{z}^2}} d\hat{z} = \frac{4\gamma}{\pi} \frac{T}{m} \tag{33}$$

However, for the resonant case, denoting $\hat{z} = 8(m-1)\gamma z^3$, a simple expansion yields very different expression for the heat flux:

$$J \approx \frac{T}{3\pi m} \int_0^\infty \frac{d\hat{z}}{1+\hat{z}^2} = \frac{T}{6m} \tag{34}$$

For the resonant case, it is somewhat surprising that the heat flux in the basic approximation does not depend on the friction in this limit case. For comparison, we explored the similar limit in $\beta - FPU$, and Lennard-Jones chains with nearest-neighbor interaction. The $\beta - FPU$ potential and Lennard-Jones are given by,

$$V(u) = \frac{1}{2}u^2 + \frac{\beta}{4}u^4 \; ; V(u) = 4\bar{\varepsilon}\left[\left(\frac{\bar{\sigma}}{u}\right)^6 - \frac{1}{2}\right]^2 \tag{35}$$

where $\beta, \bar{\varepsilon}$ and $\bar{\sigma}$ are constants.

As one can expect, the nonlinearity has significant effect. Still, for constant $\gamma$, it is possible to observe that the data for both nonlinear models collapse on the curves somewhat similar to the expressions (33) and (34) (see Fig. 5 and Fig. 6).

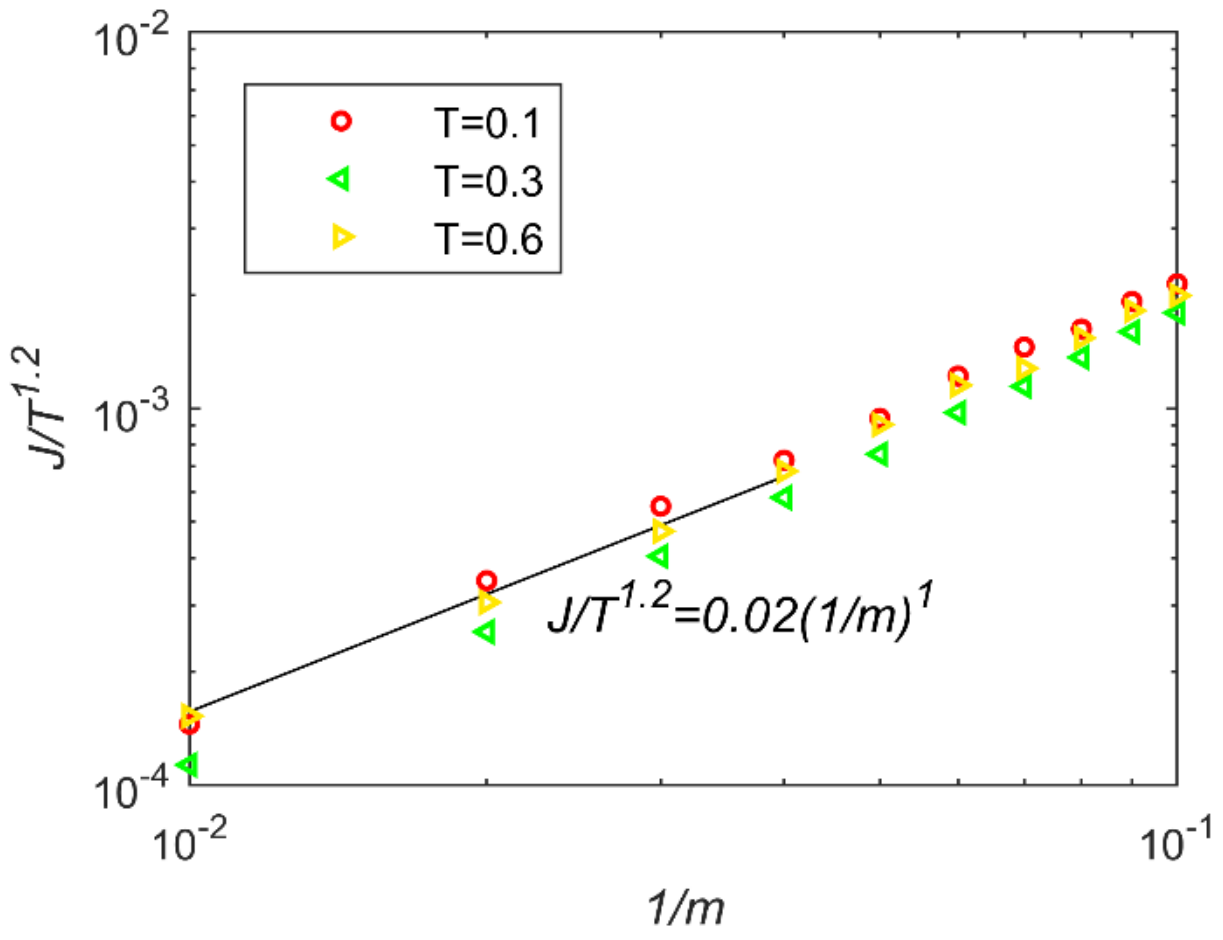

FIG. 5. (Color online) Heat flux variation with the chain temperature and defect mass for $\beta-FPU$ chain at $\gamma \ll 1, |m-1|\gamma \gg 1$. Here $N=501, \beta=0.1, \gamma=0.1$

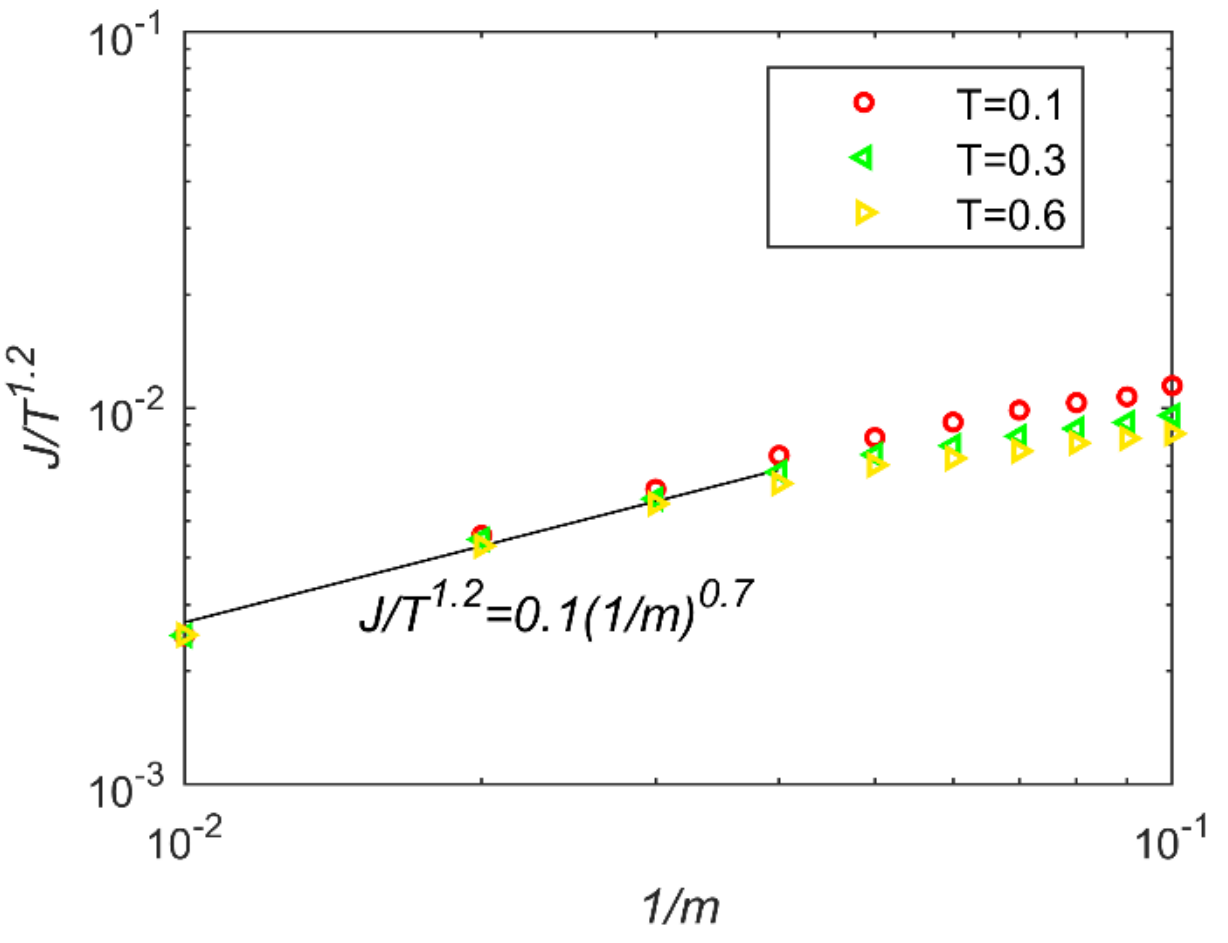

FIG. 6. (Color online) Heat flux variation with the chain temperature and defect mass for Lennard-Jones chain at $m \gg 1, \gamma \ll 1, |m-1|\gamma \gg 1$. Here $N=501, \gamma=0.1, \bar{\varepsilon}=1/72, \bar{\sigma}=2^{-1/6}$

One can conjecture that in the case of very heavy inclusion only long-wave phonons can significantly contribute to the heat flux through the defect, and thus the asymptotic behavior for linear and weakly nonlinear regimes is somewhat similar. The exponents from Figures 5 and 6 require more detailed exploration.

## 4. Temperature distribution at and near the defect. Thermal "dip" in linear and nonlinear models.

In previous sections global properties of the heat flux and temperature drop were addressed. Here we consider the details of temperature distribution at the defect site and nearby. The detailed results are presented for the resonant case; the non-resonant case exhibits very similar local behavior of the thermal profile.

According to (5, 6) the local temperature at each chain site beyond the defect is expressed as:

$$T_{n<0}=\int_0^2 \omega^2 \left|\alpha_1 \exp(inq)+\beta_1 \exp(-inq)\right|^2 d\omega$$
$$T_{n>0}=\int_0^2 \omega^2 \left|\alpha_2 \exp(-inq)+\beta_2 \exp(inq)\right|^2 d\omega \tag{36}$$

After expanding the inner part of the integral, one obtains

$$\left|\alpha_1 \exp(inq)+\beta_1 \exp(-inq)\right|^2=|\alpha_1|^2+|\beta_1|^2+\alpha_1\beta_1^*\exp(2inq)+\alpha_1^*\beta_1\exp(-2inq)$$
$$\left|\alpha_2 \exp(-inq)+\beta_2 \exp(inq)\right|^2=|\alpha_2|^2+|\beta_2|^2+\alpha_2\beta_2^*\exp(-2inq)+\alpha_2^*\beta_2\exp(2inq) \tag{37}$$

Straightforward simplification of (36) yields the following set of equations

$$T_{n<0}=\begin{cases}\frac{T}{\pi}\int_0^1\left(\begin{array}{l}\frac{\left(1+4\gamma^2z^2\right)^2+128(m-1)^2\gamma^2z^6}{\left(\left(1+4\gamma^2z^2\right)^2+64(m-1)^2\gamma^2z^6\right)\sqrt{1-z^2}}-\\-\frac{\sqrt{\left(1-z^2\right)+(m-1)^2z^2}\,128(m-1)\gamma^2z^5\sin\left(\theta_1-2nq\right)}{\left(\left(1+4\gamma^2z^2\right)^2+64(m-1)^2\gamma^2z^6\right)\sqrt{1-z^2}}+\\+\frac{|m-1|z\left(\left(1+4\gamma^2z^2\right)^2-64\gamma^2z^4\left(1-z^2\right)\right)\sin\left(\theta_1-2nq\right)}{\sqrt{\left(1-z^2\right)+(m-1)^2z^2}\left(\left(1+4\gamma^2z^2\right)^2+64(m-1)^2\gamma^2z^6\right)\sqrt{1-z^2}}\end{array}\right)dz, \text{ if } m<1\\ \frac{T}{\pi}\int_0^1\left(\begin{array}{l}\frac{\left(1+4\gamma^2z^2\right)^2+128(m-1)^2\gamma^2z^6}{\left(\left(1+4\gamma^2z^2\right)^2+64(m-1)^2\gamma^2z^6\right)\sqrt{1-z^2}}-\\-\frac{\sqrt{\left(1-z^2\right)+(m-1)^2z^2}\,128(m-1)\gamma^2z^5\sin\left(\theta_1-2nq\right)}{\left(\left(1+4\gamma^2z^2\right)^2+64(m-1)^2\gamma^2z^6\right)\sqrt{1-z^2}}-\\-\frac{|m-1|z\left(\left(1+4\gamma^2z^2\right)^2-64\gamma^2z^4\left(1-z^2\right)\right)\sin\left(\theta_1-2nq\right)}{\sqrt{\left(1-z^2\right)+(m-1)^2z^2}\left(\left(1+4\gamma^2z^2\right)^2+64(m-1)^2\gamma^2z^6\right)\sqrt{1-z^2}}\end{array}\right)dz, \text{ if } m>1\end{cases}$$

$$T_{n>0}=\begin{cases}\frac{T}{\pi}\int_0^1\left(\begin{array}{l}\frac{\left(1+4\gamma^2z^2\right)^2}{\left(\left(1+4\gamma^2z^2\right)^2+64(m-1)^2\gamma^2z^6\right)\sqrt{1-z^2}}+\\+\frac{|m-1|z\left(\left(1+4\gamma^2z^2\right)^2-64\gamma^2z^4\left(1-z^2\right)\right)\sin\left(\theta_1+2nq\right)}{\sqrt{\left(1-z^2\right)+(m-1)^2z^2}\left(\left(1+4\gamma^2z^2\right)^2+64(m-1)^2\gamma^2z^6\right)\sqrt{1-z^2}}\end{array}\right)dz, \text{ if } m<1\\ \frac{T}{\pi}\int_0^1\left(\begin{array}{l}\frac{\left(1+4\gamma^2z^2\right)^2}{\left(\left(1+4\gamma^2z^2\right)^2+64(m-1)^2\gamma^2z^6\right)\sqrt{1-z^2}}-\\-\frac{|m-1|z\left(\left(1+4\gamma^2z^2\right)^2-64\gamma^2z^4\left(1-z^2\right)\right)\sin\left(\theta_1+2nq\right)}{\sqrt{\left(1-z^2\right)+(m-1)^2z^2}\left(\left(1+4\gamma^2z^2\right)^2+64(m-1)^2\gamma^2z^6\right)\sqrt{1-z^2}}\end{array}\right)dz, \text{ if } m>1\end{cases} \quad (38)$$

Here $\theta_1=\arcsin\left(\frac{(m-1)z}{\sqrt{\left(1-z^2\right)+(m-1)^2z^2}}\right)$ and $q=2\arcsin(z)$ (39)

At the defect site itself, the temperature is expressed as

$$T_{n=0} = \begin{cases} \dfrac{mT}{\pi}\displaystyle\int_0^1 \dfrac{\left(1+4\gamma^2z^2\right)^2}{\left(\left(1+4\gamma^2z^2\right)^2+64\left(m-1\right)^2\gamma^2z^6\right)\sqrt{1-z^2}} + \\ + \dfrac{|m-1|\left(m-1\right)\left(\left(1+4\gamma^2z^2\right)^2-64\gamma^2z^4\left(1-z^2\right)\right)z^2}{\left(\left(1-z^2\right)+\left(m-1\right)^2z^2\right)\left(\left(1+4\gamma^2z^2\right)^2+64\left(m-1\right)^2\gamma^2z^6\right)\sqrt{1-z^2}}dz, \text{ if } m<1 \\ \dfrac{mT}{\pi}\displaystyle\int_0^1 \dfrac{\left(1+4\gamma^2z^2\right)^2}{\left(\left(1+4\gamma^2z^2\right)^2+64\left(m-1\right)^2\gamma^2z^6\right)\sqrt{1-z^2}} - \\ - \dfrac{|m-1|\left(m-1\right)\left(\left(1+4\gamma^2z^2\right)^2-64\gamma^2z^4\left(1-z^2\right)\right)z^2}{\left(\left(1-z^2\right)+\left(m-1\right)^2z^2\right)\left(\left(1+4\gamma^2z^2\right)^2+64\left(m-1\right)^2\gamma^2z^6\right)\sqrt{1-z^2}}dz, \text{ if } m>1 \end{cases} \quad (40)$$

For the case of very small mass detuning, $|m-1| \ll 1$, one obtains the following expression for the temperature at the defect:.

$$T_{n=0} = \frac{mT}{\pi}\int_0^1 \left( \frac{1}{\sqrt{1-z^2}} + \frac{|m-1|\left(m-1\right)z^2}{\left(\left(m-1\right)^2-1\right)\left(z^2+\dfrac{1}{\left(m-1\right)^2-1}\right)\sqrt{1-z^2}} \right) dz, \text{ if } 1-m \ll 1 \quad (41)$$

These expressions simplify to

$$T_{n=0} = \begin{cases} T\left(m-\tfrac{1}{2}\right), \text{ if } m<1 \\ T/2, \text{ if } m>1 \end{cases} \quad (42)$$

This peculiar non-analytic dependence of the defect temperature on $m$ calls for closer exploration. In Fig. 7, $T_1$, $T_2$ and $T_{n=0}$ are presented versus the defect mass $m$.

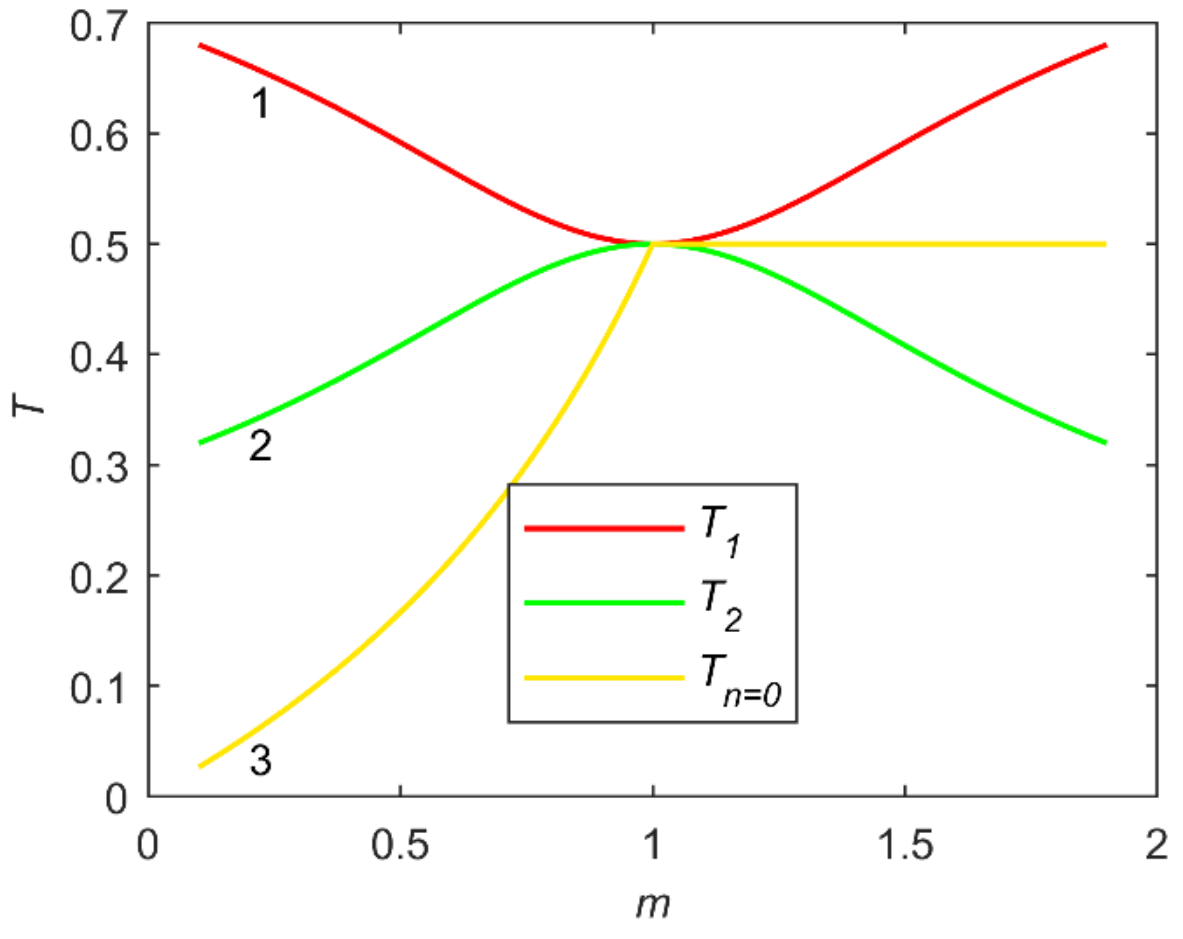

FIG. 7. (Color online) $T_1$ (marker 1), $T_2$ (marker 2) and $T_{n=0}$ (marker 3) versus $m$ ($T=1, \gamma=1$). $T_1$ and $T_2$ are symmetric to both sides of $m=1$, whereas $T_{n=0}$ is not and it is less than $T_2$ (thermal dip) for $m<1$.

It is clear from Fig. 7 that for the case $m<1$ (lighter isotopic impurity), the temperature of the defect is less than $T_2$, therefore the temperature profile exhibits a thermal "dip" at the light defect. Such dips are apparently inconsistent with the Second Law of thermodynamics, since the heat flows from "cold" defect to "hotter" part of the chain [24, 25]. To explain the appearance of the dip, one notes that for $0<m<1$, the spectrum of the chain with defect contains a localized mode in the attenuation zone. For chain long enough, the thermostats cannot excite this localized mode, thus causing the dip. In the linear chain the modes do not exchange energy, therefore there are no mechanisms for achieving the thermal equilibrium. By this reason, the "dip" in the linear chain is persistent. To illustrate this point, we simulate the modal energy distribution [29] for rather short chain with $N=51$ and light isotopic defect placed at the middle or close to one of the thermostats. The results are presented in Fig. 8.

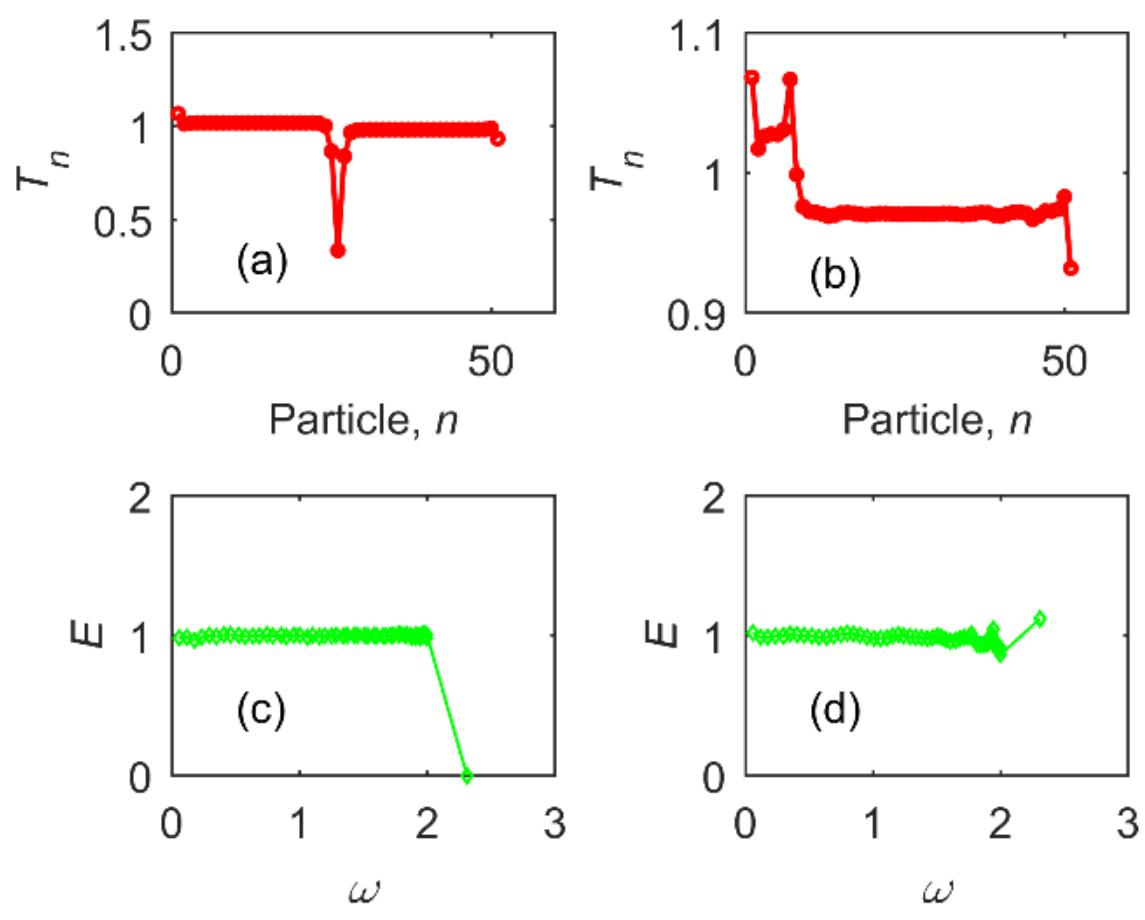

FIG. 8. (Color online) The temperature profile ((a) and (b)) and the energy spectrum ((c) and (d)). The isotopic defect is placed at the middle of the chain at $n=26$ ((a) and (c)) and near the hot thermostat at $n=7$ ((b) and (d)), $N=51, T_+=1.1, T_-=0.9, \gamma=1, m=0.5$.

In Fig. 8, when the defect is at the middle of the chain, the localized mode, is almost not excited. As the defect is closer to the thermostat, a peak replaces the dip. In this case, the localized mode is strongly excited. This latter case is not described by the previous analytic treatment, since the transition to the infinite chain length is irrelevant here.

The aforementioned anomalies of the thermal profile are caused by peculiarities of the frequency spectrum and by lack of interaction between the modes in the linear system. Thus, one can expect that the nonlinearity will remove, or, at least, significantly modify the temperature distribution at the defect. To check this assumption, we simulated the $\beta - FPU$ chain [30] with the light isotopic defect. Time dynamics of the thermal profile in this system is presented in Fig. 9 and $T_{\pm}$ represents the hot and cold thermostats at the boundaries.

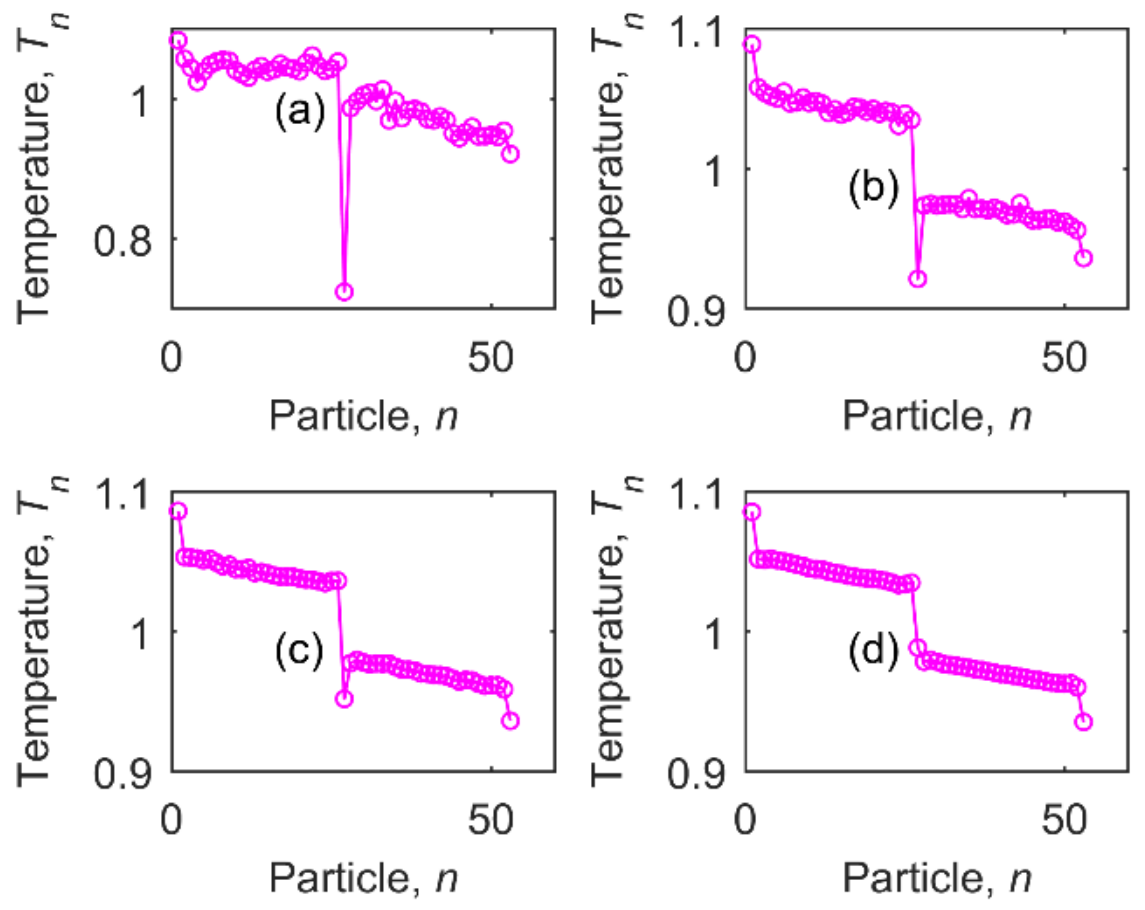

FIG. 9. (Color online) Destruction of the dip in the temperature profile in the $\beta - FPU$ chain. Each plot shows temperature profile after $10^6$ time steps (a), $10^7$ time steps (b), $10^8$ time steps (c), and $10^9$ time steps (d) respectively, $N = 51, T_{\pm} = 1 \pm 0.1, \gamma = 1, m = 0.5, \beta = 0.1$.

Fig. 9 demonstrates that at relatively short time scale the dip is observed even in the nonlinear chain. As the simulation evolves, the dip gets gradually destroyed and the defect particle temperature gets lifted close to the average chain temperature. Simple explanation is that as the nonlinearity is relatively small, the initial thermal profile is established through approximately linear dynamics (see also Refs. [31,32]). Nonlinearity reveals itself at higher time scale, related to the temperature. For lower temperatures, the process of the dip elimination is expected to take more time. In the FPU chain one can discuss two possible basic mechanisms of the dip elimination - resonance of the renormalized waves and excitation of the defect by mobile discrete breathers, existing in the frequency range of the localized mode [26,27]. Detailed analysis of these mechanisms is beyond the scope of this paper. We report some numeric observations related to the dip elimination in two different settings – in the chain with only one thermostat (equilibrium simulation), and with temperature gradient and nonzero heat flux (non-equilibrium). In both settings, initially the dip exists, and is gradually eliminated; however, the dynamics of this elimination demonstrates substantial differences.

The easy way to observe the breather is to observe the spatiotemporal profile for the frequency range of interest. Such picture is obtained by filtering out all unnecessary frequencies from the time history of the system. We consider a one-dimensional chain of length $N$ with light isotopic defect at the middle. As it was mentioned above, two systems are considered: the non-equilibrium system with $N = 51$, fixed boundary and thermostats at each boundary, and the equilibrium system with $N = 50$, periodic boundary conditions and one thermostat in the closed chain. The Langevin thermostat with a coupling friction $\gamma = 1$, temperature $T = 1$ is considered, according to (4). To improve the visibility, rather light defect with $m = 0.05$ is taken. The system is run initially with $\beta = 0$ (linear, see (35)) for $10^7$ steps, and subsequently the nonlinearity is switched on with $\beta = 0.1$. Immediately from the first step, for all particles, the particle-position history is recorded for $10^5$ steps. Then, Fourier transform is taken for these data and all frequency components other than in the range of linear localized mode are filtered out. Here, the linear frequency of the localized mode is equal to 6.4 and the filtering range is $6 \le \omega \le 7$. After filtering, the inverse Fourier transform is taken.

In Fig. 10, the interaction between a moving breather and the defect particle is clearly observed for the non-equilibrium simulation. The energy of the defect particle increases after the interaction with the breather.

In the case of single thermostat (equilibrium simulation) the breather propagation has not been observed. We can conjecture that in the conditions of directed heat flux the probability of excitation of the mobile breather is higher than in the conditions of stationary temperature distribution, without the directed heat flux. This difference is reflected in different time required to destroy the dip in these two settings, as presented below.

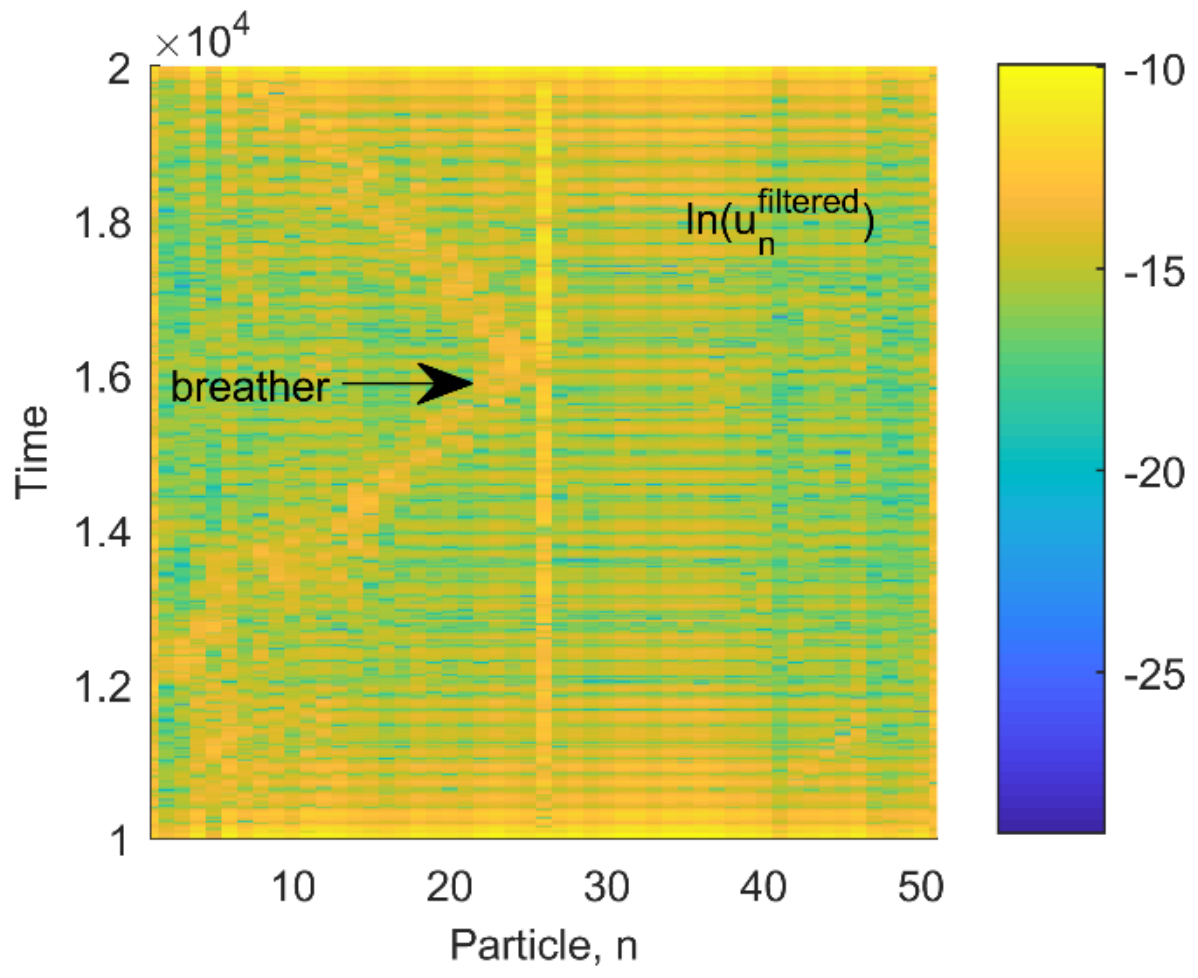

FIG. 10. (Color online) Breather starts from the thermostat and excites the defect particle in the non-equilibrium system. Frequency filtering range $6 \le \omega \le 7$, $m = 0.05, T_{\pm} = 1 \pm 0.1, \beta = 0.1,$.

To quantify the dynamics of the dip elimination, the normalized temperature of the defect particle is defined as follows:

$$T^* = \frac{T_{Ave} - T_{Def}}{\left(T_{Ave} - T_{Def}\right)_{\max}} \tag{43}$$

$T_{Ave}$ is the average temperature of the system and $T_{Def}$ is the temperature of the defect particle, as functions of time. First, the non-equilibrium case is considered. To suppress the fluctuations, the average of 1800 realizations is taken. The excitation profile is studied for 20 different cases of the defect mass. The latter were chosen to yield particular linear localized frequencies, selected between $2.1 \le \omega_{loc} \le 4$, with 0.1 increment. The time series for $T^*$ in the case of the non-equilibrium simulation are presented in Fig. 11 and for the equilibrium simulation - in Fig. 12. In general, these time series have three regions. The first region is almost horizontal, where the nonlinearity almost does not reveal itself. Then one observes the excitation region where the defect particle receives energy as the time progresses. In the third region, the defect particle reaches a steady state at which $T^*$ starts fluctuating around positive and negative values. One can observe that in the non-equilibrium case the temperature dip is removed at shorter time scale. It is possible to conjecture that the process is expedited by interaction of the defect with mobile breathers created at the thermostats. For the equilibrium case, the excitation profile shows linear trend with the evolution time for all simulated values of defect masses (cf. Fig. 12) and temperatures (cf. Fig. 13).

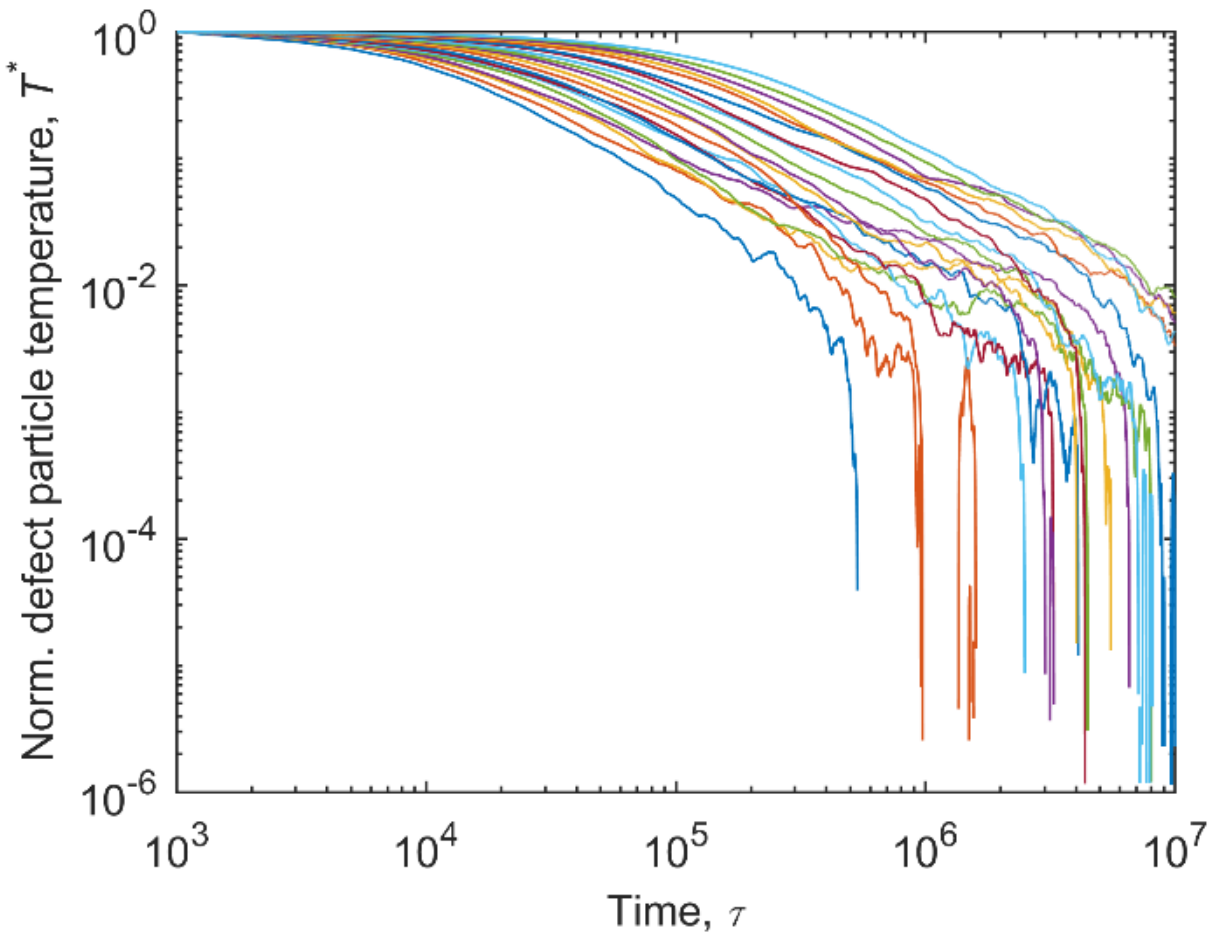

FIG. 11. (Color online) Normalized defect temperature for the non-equilibrium simulations, for 20 different defect masses, $m = 0.70 - 0.13, T_{\pm} = 1 \pm 0.1, \beta = 0.1$

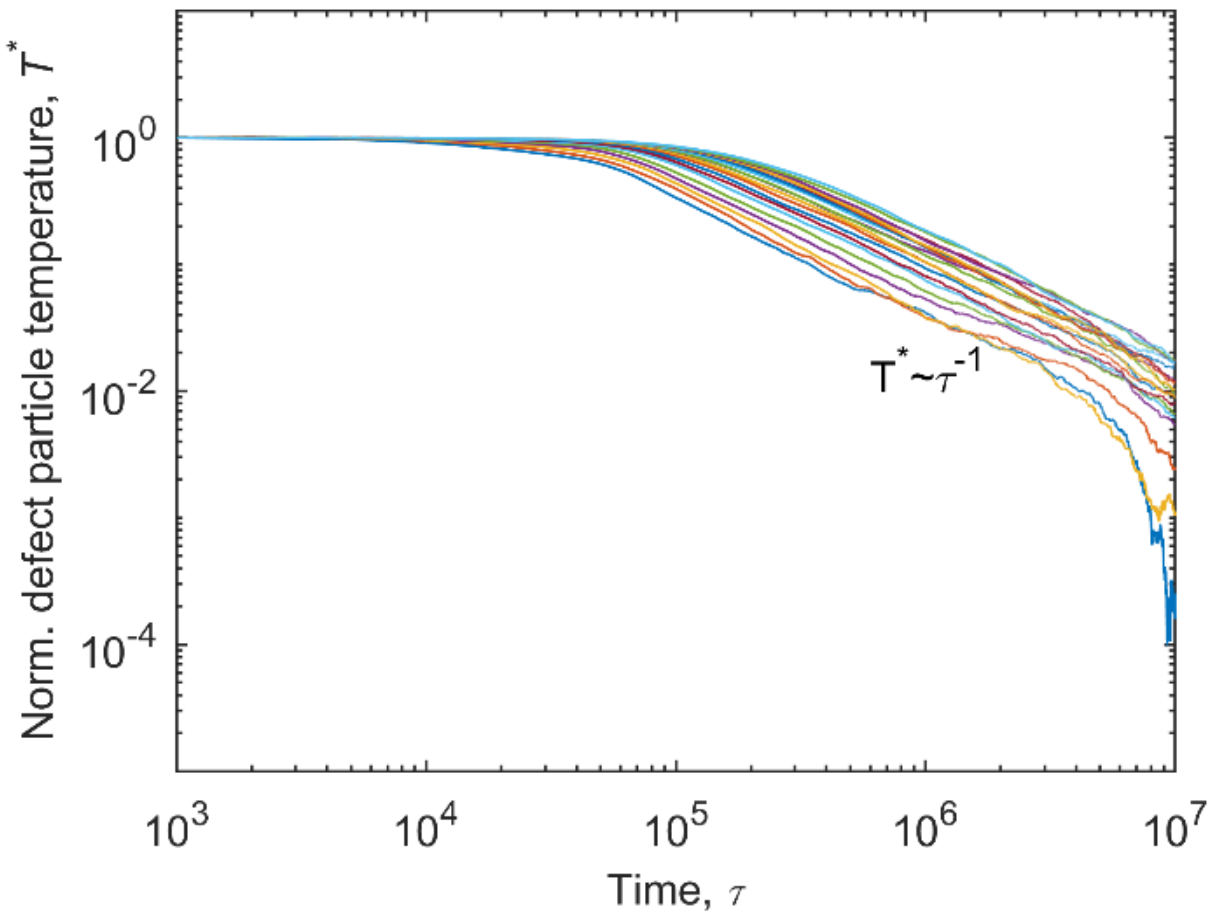

FIG. 12. (Color online) Normalized defect temperature for the equilibrium simulations, for 20 different defect masses, $m = 0.70 - 0.13, T_{\pm} = 1 \pm 0.1, \beta = 0.1$

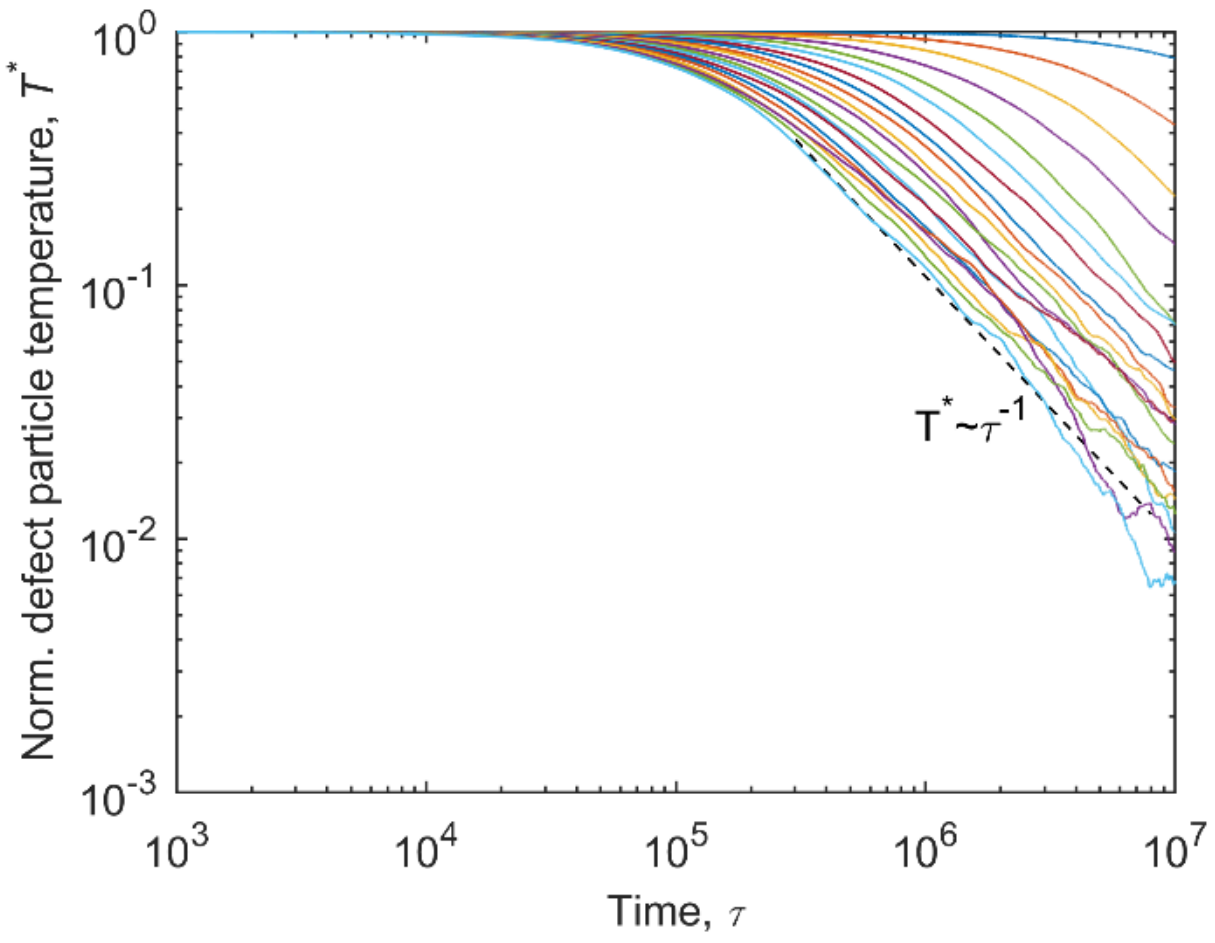

FIG. 13. (Color online) Normalized defect temperature for the equilibrium simulations in temperature interval $T = 0.1 - 2$. As in Fig. 12, the excitation profile follows linear trend. $m = 0.05, \beta = 0.1$

It is well known [17] that in linear chains the temperature profile near the thermostats exponentially achieves the average values. This transient occurs due to the waves irradiated by the thermostat, with frequencies belonging to the attenuation zone of the chain. In the present case, in the vicinity of the isolated defect, one also observes the transient to the constant temperature (see, e.g. Fig. 8), despite the fact that in the case of very long chain all frequencies active in this region belong to the propagation zone.

To explain this peculiarity of the temperature profile around the defect site, one defines

$$T_{n<0} = T_1 + \Delta T_{n<0}; T_{n>0} = T_2 + \Delta T_{n>0}, \tag{44}$$

Then, with account of (38), it is easy to obtain

$$\Delta T_{n<0} = \begin{cases} \dfrac{T}{\pi}\displaystyle\int_0^1 \left( \begin{array}{l} -\dfrac{\sqrt{(1-z^2)+(m-1)^2 z^2}\,128(m-1)\gamma^2 z^5 \sin(\theta_1 - 2nq)}{\left((1+4\gamma^2 z^2)^2 + 64(m-1)^2\gamma^2 z^6\right)\sqrt{1-z^2}} + \\ + \dfrac{|m-1| z\left((1+4\gamma^2 z^2)^2 - 64\gamma^2 z^4 (1-z^2)\right)\sin(\theta_1 - 2nq)}{\sqrt{(1-z^2)+(m-1)^2 z^2}\,((1+4\gamma^2 z^2)^2 + 64(m-1)^2\gamma^2 z^6)\sqrt{1-z^2}} \end{array} \right) dz, \text{ if } m<1 \\ \dfrac{T}{\pi}\displaystyle\int_0^1 \left( \begin{array}{l} -\dfrac{\sqrt{(1-z^2)+(m-1)^2 z^2}\,128(m-1)\gamma^2 z^5 \sin(\theta_1 - 2nq)}{\left((1+4\gamma^2 z^2)^2 + 64(m-1)^2\gamma^2 z^6\right)\sqrt{1-z^2}} - \\ - \dfrac{|m-1| z\left((1+4\gamma^2 z^2)^2 - 64\gamma^2 z^4 (1-z^2)\right)\sin(\theta_1 - 2nq)}{\sqrt{(1-z^2)+(m-1)^2 z^2}\left((1+4\gamma^2 z^2)^2 + 64(m-1)^2\gamma^2 z^6\right)\sqrt{1-z^2}} \end{array} \right) dz, \text{ if } m>1 \end{cases}$$

$$\Delta T_{n>0} = \begin{cases} \dfrac{T}{\pi}\displaystyle\int_0^1 \left( + \dfrac{|m-1| z\left((1+4\gamma^2 z^2)^2 - 64\gamma^2 z^4 (1-z^2)\right)\sin(\theta_1 + 2nq)}{\sqrt{(1-z^2)+(m-1)^2 z^2}\left((1+4\gamma^2 z^2)^2 + 64(m-1)^2\gamma^2 z^6\right)\sqrt{1-z^2}} \right) dz, \text{ if } m<1 \\ \dfrac{T}{\pi}\displaystyle\int_0^1 \left( - \dfrac{|m-1| z\left((1+4\gamma^2 z^2)^2 - 64\gamma^2 z^4 (1-z^2)\right)\sin(\theta_1 + 2nq)}{\sqrt{(1-z^2)+(m-1)^2 z^2}\left((1+4\gamma^2 z^2)^2 + 64(m-1)^2\gamma^2 z^6\right)\sqrt{1-z^2}} \right) dz, \text{ if } m>1 \end{cases} \tag{45}$$

We consider the case $n > 0, m < 1$, with trivial generalization for other cases. To proceed with the analytic evaluation, one assumes $|m-1| \ll 1, \gamma \ll 1$ and obtains

$$\Delta T_n \approx \frac{T}{\pi}\int_0^1 \left( \begin{array}{l} \dfrac{|m-1|\varepsilon z^2}{\left(\left(1-z^2\right)+\left(m-1\right)^2 z^2\right)\sqrt{1-z^2}}\cos\left(4n\arcsin\left(z\right)\right) \\ +\dfrac{|m-1|z}{\left(1-z^2\right)+\left(m-1\right)^2 z^2}\sin\left(4n\arcsin\left(z\right)\right) \end{array} \right) dz \tag{46}$$

We denote $\varepsilon = m-1$ and $z = \sin\left(q/2\right)$, and obtain.

$$\Delta T_n \approx \frac{T}{2\pi}\int_0^{\pi} \left( \begin{array}{l} \dfrac{|\varepsilon|\varepsilon\sin^2\left(q/2\right)}{\left(1-\sin^2\left(q/2\right)\right)+\varepsilon^2\sin^2\left(q/2\right)}\cos\left(2nq\right)+ \\ +\dfrac{|\varepsilon|\sin\left(q/2\right)\cos\left(q/2\right)}{\left(1-\sin^2\left(q/2\right)\right)+\varepsilon^2\sin^2\left(q/2\right)}\sin\left(2nq\right) \end{array} \right) dq$$

$$= \frac{T}{4\pi\left(1+\varepsilon^2\right)}\int_0^{2\pi} \left( \begin{array}{l} \dfrac{|\varepsilon|\varepsilon\left(1-\cos q\right)\cos\left(2nq\right)}{1+\upsilon\cos q}+ \\ +\dfrac{|\varepsilon|\sin q\sin\left(2nq\right)}{1+\upsilon\cos q} \end{array} \right) dq \tag{47}$$

Here $\upsilon = \dfrac{1-\varepsilon^2}{1+\varepsilon^2}$. Then, Eq. (47) is transferred to the complex plane by substituting

$\kappa = \exp\left(i\theta\right)$, and then reduced to the form

$$\Delta T_n \approx \frac{-iT}{8\pi\upsilon\left(1+\varepsilon^2\right)}\oint_{|\kappa|=1} \left( \begin{array}{l} \mathrm{Re}\left(\dfrac{|\varepsilon|\varepsilon(2\kappa^{2n}-\kappa^{2n+1}-\kappa^{2n-1}}{\kappa^2+2/\upsilon\,\kappa+1}\right)+ \\ +\mathrm{Im}\left(\dfrac{|\varepsilon|\left(\kappa^{2n-1}-\kappa^{2n+1}\right)}{\kappa^2+2/\upsilon\,\kappa+1}\right) \end{array} \right) d\kappa \tag{48}$$

The poles in (48) are $-\dfrac{1}{\upsilon}\pm\sqrt{\dfrac{1}{\upsilon^2}-1}$ and the one which is inside the unit circle is

$-\dfrac{1}{\upsilon}+\sqrt{\dfrac{1}{\upsilon^2}-1}$, then the expression simplifies to,

$$
\Delta T_n \approx \frac{T}{4\sqrt{1-\upsilon^2}\left(1+\varepsilon^2\right)}\times
$$

$$
\times\left(\begin{array}{l} |\varepsilon|\varepsilon\left(2\left(-\frac{1}{\upsilon}+\sqrt{\frac{1}{\upsilon^2}-1}\right)^{2n}-\left(-\frac{1}{\upsilon}+\sqrt{\frac{1}{\upsilon^2}-1}\right)^{2n+1}-\left(-\frac{1}{\upsilon}+\sqrt{\frac{1}{\upsilon^2}-1}\right)^{2n-1}\right)+ \\ +|\varepsilon|\left(\left(-\frac{1}{\upsilon}+\sqrt{\frac{1}{\upsilon^2}-1}\right)^{2n-1}-\left(-\frac{1}{\upsilon}+\sqrt{\frac{1}{\upsilon^2}-1}\right)^{2n+1}\right) \end{array}\right) \tag{49}
$$

The result means that the near - field of the defect follows staggering exponential convergence to the average value. Interestingly, the exponent is governed solely by the mass of defect. In Fig. 14, the exponential decay to the average temperature is illustrated by the numerical integration of (45).

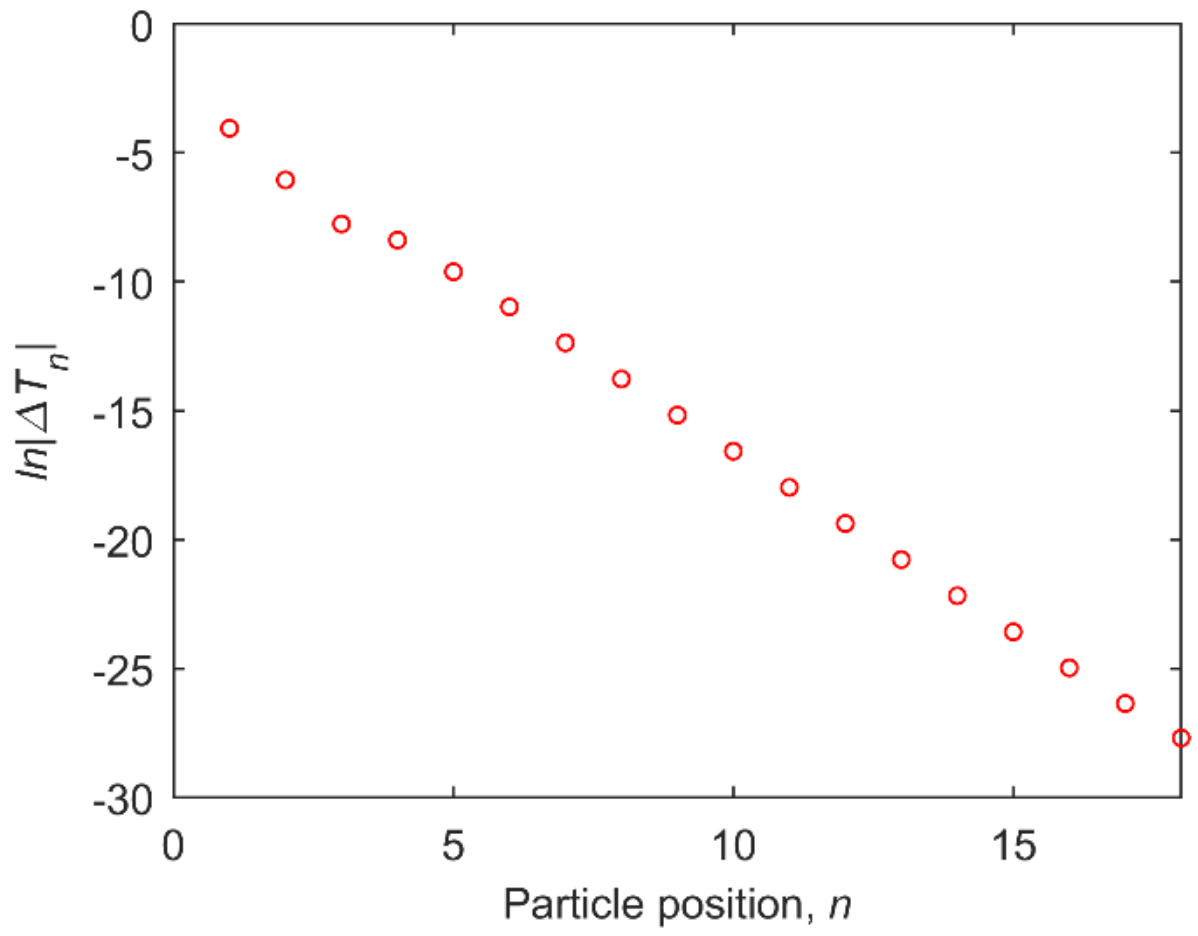

FIG. 14. (Color online) The absolute value of $\Delta T_n$ versus particle number, semi-logarithmic scale. $T=1, \gamma=1.2, m=1.6$.

## 5. Concluding remarks.

For linear chain with arbitrary local conservative LTI defect, one can derive the explicit exact expressions for the heat flux and the temperature drop. The expressions are generic, and illustrated by a simple case of the isolated isotopic defect. For more complex inclusions, the expression for the transfer matrix will also become more complicated, but still will be at least in principle derivable by linear algebra.

The Kapitza resistance in the linear chain, albeit finite, is not a local property of the defect and its close surrounding. This lack of locality is expressed in two main ways. The first one is the dependence of the results on exact way of approaching the thermodynamic limit. In fact, the universal limit of the infinite system size for the resistance does not exist. The analysis reveals clear distinction between the resonant and non-resonant cases, both in the observed values of the resistance, and in the asymptotic dependence on the mass mismatch. Besides, the resistance value strongly depends on the thermostat characteristics.

The analysis presented in the paper is suitable for one-dimensional setting. 2D and 3D settings offer much more possibilities for shapes of the inclusions; therefore, the problem of the thermal resistance should include detailed specification of the geometry. At the same time, our treatment is based on ideas and methods of the linear response theory that are not bounded to one dimension. So, in principle, the solution for multi-dimensional model may be formulated based on the same ideas. From the other side, the transition to the thermodynamic limit will be even more tricky, and the result will depend on it. Besides, one can also expect the dependence of the resistance on the thermostat properties, similarly to the 1D case.

Another salient feature of the linear model is the thermal dip at the light defect site, related to insufficient excitation of the localized mode by the thermostat, and by lack of the intermodal interactions. When the nonlinearity is switched on, the dip gradually disappears, thus repairing the apparent violation of the Second Law of thermodynamics. Of course, it does not mean that the thermal resistance in the "repaired" nonlinear system will be normal in all aspects [23]. Time dynamics of the dip removal is different in the equilibrium and non-equilibrium simulations. We conjecture that the difference is related to creation of propagating breathers in the non-equilibrium system, that can interact with the defect and further excite it. Intrinsic dynamics of this process of the dip removal requires further exploration.

It is important to mention that the dip appears in the temperature profile even for the nonlinear system, although for relatively short time. This time scale is governed by the system size and the defect mass, as the linear substructure is non-dimensional and normalized. The time scale necessary to reveal the nonlinearity strongly depends on the temperature. Thus, if the temperature is low enough, the initial response of the system to the imposed thermal gradient will be governed by the linear dynamics. In other terms, the linear dynamics may be useful to understand the short-time response of more realistic nonlinear models. This latter conclusion points to the fundamental difference between the steady-state thermal resistance and the short-time response. For the steady state, one encounters qualitative differences between the linear and nonlinear models [23], e.g. the strong size dependence of the resistance.

**Acknowledgment**

The authors are very grateful to Israel Science Foundation (Grant No. 1696/17) for financial support.

## References

[1] P. L. Kapitza, The study of heat transfer in Helium II, J. Phys. USSR **4**, 181 (1941).
[2] G. L. Pollack, *Kapitza Resistance*, Rev. Mod. Phys. **41**, 48 (1969).
[3] N. S. Snyder, *Heat Transport through Helium II: Kapitza Conductance*, Cryogenics **10**, 89 (1970).
[4] L. J. Challis, *Kapitza Resistance and Acoustic Transmission across Boundaries at High Frequencies*, J. Phys. C Solid State Phys. **7**, 481 (1974).
[5] E. T. Swartz and R. O. Pohl, *Thermal Boundary Resistance*, Rev. Mod. Phys. **61**, 605 (1989).
[6] I. M. Khalatnikov, *An Introduction To The Theory Of Superfluidity* (CRC Press, 2018).
[7] W. A. Little, *The Transport of Heat between Dissimilar Solids at Low Temperatures*, Can. J. Phys. **37**, 334 (1959).
[8] S. Simons, *On the Thermal Contact Resistance between Insulators*, J. Phys. C Solid State Phys. **7**, 4048 (1974).
[9] R. E. Peterson and A. C. Anderson, *The Kapitza Thermal Boundary Resistance*, J. Low Temp. Phys. **11**, 639 (1973).
[10] P. Erdös and S. B. Haley, *Low-Temperature Thermal Conductivity of Impure Insulators*, Phys. Rev. **184**, 951 (1969).
[11] H. Budd and J. Vannimenus, *Thermal Boundary Resistance*, Phys. Rev. Lett. **26**, 1637 (1971).
[12] W. M. Saslow, *Kapitza Conductance, Temperature Gradients, and Solutions to the Boltzmann Equation*, Phys. Rev. B **11**, 2544 (1975).
[13] T. J. Sluckin, *A Simple Model for the Kapitza Conductance*, Phys. Lett. A **53**, 390 (1975).
[14] Ch. Steinbrüchel, *The Scattering of Phonons of Arbitrary Wavelength at a Solid-Solid Interface: Model Calculation and Applications*, Z. Für Phys. B Condens. Matter Quanta **24**, 293 (1976).
[15] M. E. Lumpkin, W. M. Saslow, and W. M. Visscher, *One-Dimensional Kapitza Conductance: Comparison of the Phonon Mismatch Theory with Computer Experiments*, Phys. Rev. B **17**, 4295 (1978).
[16] E. T. Swartz and R. O. Pohl, *Thermal Resistance at Interfaces*, Appl. Phys. Lett. **51**, 2200 (1987).
[17] Z. Rieder, J. L. Lebowitz, and E. Lieb, *Properties of a Harmonic Crystal in a Stationary Nonequilibrium State*, J. Math. Phys. **8**, 1073 (1967).
[18] A. Dhar and D. Roy, *Heat Transport in Harmonic Lattices*, J. Stat. Phys. **125**, 801 (2006).
[19] A. Casher and J. L. Lebowitz, *Heat Flow in Regular and Disordered Harmonic Chains*, J. Math. Phys. **12**, 1701 (1971).
[20] V. Kannan, A. Dhar, and J. L. Lebowitz, *Nonequilibrium Stationary State of a Harmonic Crystal with Alternating Masses*, Phys. Rev. E **85**, 041118 (2012).
[21] T. Mai, A. Dhar, and O. Narayan, *Equilibration and Universal Heat Conduction in Fermi-Pasta-Ulam Chains*, Phys. Rev. Lett. **98**, 184301 (2007).
[22] V. Kannan, Heat Conduction in Low Dimensional Lattice Systems, Rutgers University - Graduate School - New Brunswick, 2013.
[23] J. Paul and O. V. Gendelman, *Kapitza Resistance in Basic Chain Models with Isolated Defects*, Phys. Lett. A **384**, 126220 (2020).
[24] X. Cao and D. He, *Interfacial Thermal Conduction and Negative Temperature Jump in One-Dimensional Lattices*, Phys. Rev. E **92**, 032135 (2015).

[25] Y. Liu and D. He, *Anomalous Interfacial Temperature Profile Induced by Phonon Localization*, Phys. Rev. E **96**, 062119 (2017).
[26] B. Gershgorin, Y. V. Lvov, and D. Cai, *Renormalized Waves and Discrete Breathers in β - Fermi-Pasta-Ulam Chains*, Phys. Rev. Lett. **95**, 264302 (2005).
[27] N. Li, B. Li, and S. Flach, *Energy Carriers in the Fermi-Pasta-Ulam β Lattice: Solitons or Phonons?*, Phys. Rev. Lett. **105**, 054102 (2010).
[28] J. W. S. Rayleigh, *Treatise on Sound Vol II* (London: Macmillan), 1878).
[29] L. Meirovitch, *Fundamentals of Vibrations* (McGraw-Hill, Boston, Mass, 2001).
[30] E. Fermi, P. Pasta, S. Ulam, and M. Tsingou, Studies of the Nonlinear Problems, Los Alamos Scientific Lab., N. Mex., 1955.
[31] V.A. Kuzkin and A.M. Krivtsov, *Ballistic resonance and thermalization in the Fermi-Pasta-Ulam-Tsingou chain at finite temperature,* Phys. Rev. E, **101**, 042209 (2020)
[32] E.A. Korznikova, V.A. Kuzkin, A.M. Krivtsov, D. Xiong, V.A. Gani, A.A. Kudreyko and S.V. Dmitriev, *Equilibration of sinusoidal modulation of temperature in linear and nonlinear chains* , Phys. Rev. E, **102**, 062148 (2020)

## APPENDIX A: AVERAGIND IN THE LIMIT OF THE INFINITE CHAIN LENGTH.

### 1. The non-resonant case.

From Eq. (17-18) one obtains:

$$\left\langle \frac{1}{|D|^2} \right\rangle_{\substack{N_1\to\infty \\ N_2\to\infty}} = \frac{1}{4\pi^2}\int_0^{2\pi} d\varphi_1 \int_0^{2\pi} d\varphi_2 \times$$

$$\times \frac{1}{\left|\cosh x e^{i(\varphi_1+\varphi_2+\theta_1)} c_+^2 + \sinh x c_+ c_- (e^{i(\varphi_2-\varphi_1+\theta_2)} - e^{-i(\varphi_2-\varphi_1+\theta_2)}) - \cosh x e^{-i(\varphi_1+\varphi_2+\theta_1)} c_-^2\right|^2} \underset{\substack{\zeta_1=\varphi_1+\frac{\theta_1-\theta_2}{2} \\ \zeta_2=\varphi_2+\frac{\theta_1+\theta_2}{2}}}{=}$$

$$= \frac{1}{4\pi^2}\int_0^{2\pi} d\zeta_1 \int_0^{2\pi} d\zeta_2 \frac{1}{\left(Ae^{i\zeta_1} - Be^{-i\zeta_1}\right)\left(A^* e^{-i\zeta_1} - B^* e^{i\zeta_1}\right)} = \frac{1}{2\pi}\int_0^{2\pi} \frac{d\zeta_2}{\left||A|^2 - |B|^2\right|}$$

$$A = \cosh x e^{i\zeta_2} c_+^2 - \sinh x c_+ c_- e^{-i\zeta_2}, B = \cosh x e^{-\zeta_2} c_-^2 - \sinh x c_+ c_- e^{i\zeta_2}$$

(A1)

Further integration yields:

$$\frac{1}{2\pi}\int_0^{2\pi} \frac{d\zeta_2}{\left||A|^2 - |B|^2\right|} \underset{j=\exp(2i\zeta_2)}{=} \frac{-i}{2\pi\left(|c_+|^2 - |c_-|^2\right)} \oint_{|j|=1} \frac{dj}{j(Q - Pj - P^*/j)} = \frac{1}{\left(|c_+|^2 - |c_-|^2\right)\sqrt{Q^2 - 4PP^*}} \quad \text{(A2)}$$

$$Q = \left(|c_+|^2 + |c_-|^2\right)\cosh^2 x,\; P = \sinh x \cosh x c_+ c_-^*$$

Similarly, for evaluation of $T_1$ in (17) one obtains

$$\left\langle \frac{\exp(i(2\varphi_2+\theta_1+\theta_2)}{|D|^2} \right\rangle_{\substack{N_1\to\infty \\ N_2\to\infty}} = \frac{1}{4\pi^2}\int_0^{2\pi} d\zeta_1 \int_0^{2\pi} d\zeta_2 \frac{\exp(2i\zeta_2)}{\left(Ae^{i\zeta_1}-Be^{-i\zeta_1}\right)\left(A^* e^{-i\zeta_1}-B^* e^{i\zeta_1}\right)} =$$

$$= \frac{1}{2\pi}\int_0^{2\pi} \frac{\exp(2i\zeta_2)d\zeta_2}{\left||A|^2-|B|^2\right|} = \frac{-i}{2\pi\left(|c_+|^2-|c_-|^2\right)} \oint_{|j|=1} \frac{jdj}{j(Q-Pj-P^*/j)} =$$

$$= \frac{Q-\sqrt{Q^2-4PP^*}}{2P\left(|c_+|^2-|c_-|^2\right)\sqrt{Q^2-4PP^*}}$$

(A3)

Expressions (19, 28) follow from (A2, A3)

## 2. The resonant case.

We denote $2qN=\varphi$ and obtain:

$$\left\langle \frac{1}{|D|^2} \right\rangle_{N\to\infty} = \frac{1}{2\pi}\int_0^{2\pi} \frac{d\varphi}{\begin{bmatrix}(\cosh x e^{i\varphi}e^{i\theta_1}c_+^2+2i\sinh x\sin\theta_2 c_+c_- - \cosh x e^{-i\varphi}e^{-i\theta_1}c_-^2)\times \\ (\cosh x e^{-i\varphi}e^{-i\theta_1}c_+^{*2}-2i\sinh x\sin\theta_2 c_+^*c_-^* - \cosh x e^{i\varphi}e^{i\theta_1}c_-^{*2})\end{bmatrix}} = \quad \text{(A4)}$$

$$\underset{j=\exp(i\varphi)}{=} \frac{-i}{2\pi\cosh^2 x|c_+|^4}\oint_{|j|=1}\frac{jdj}{(j-j_1)(j-j_2)(jj_1^*-1)(jj_2^*-1)}$$

In (A4) $j_{1,2}$ are the roots of polynomial $P(j)=\cosh x e^{i\theta_1}c_+^2 j^2+2i\sinh x\sin\theta_2 c_+c_- j-\cosh x e^{-i\theta_1}c_-^2$:

$$j_{1,2}=\frac{c_-}{c_+}\frac{-i\sinh x\sin\theta_2\pm\sqrt{\cosh^2 x-\sinh^2 x\sin^2\theta_2}}{\cosh x\exp(i\theta_1)} \quad \text{(A5)}$$

It is easy to demonstrate that both roots lie inside or at the boundary of the unit circle:

$$\left|j_{1,2}\right|^2=\left|\frac{c_-}{c_+}\right|^2=\frac{1+\gamma^2\omega^2-2\gamma\omega\sin q}{1+\gamma^2\omega^2+2\gamma\omega\sin q}\le 1 \quad \text{(A6)}$$

Therefore, only these two roots contribute to the integral in (A1), two others lie outside the unit circle. Thus, one obtains:

$$\left\langle \frac{1}{|D|^2} \right\rangle_{N\to\infty} = \frac{1}{\cosh^2 x|c_+|^4}\left(\frac{j_1}{(j_1-j_2)\left(|j_1|^2-1\right)\left(j_1j_2^*-1\right)}+\frac{j_2}{(j_2-j_1)\left(j_2j_1^*-1\right)\left(|j_1|^2-1\right)}\right)=$$

$$=\frac{|c_+|^2+|c_-|^2}{\cosh^2 x\left(|c_+|^2-|c_-|^2\right)\left(|c_+|^4+|c_-|^4+2|c_+|^2|c_-|^2\left(1-2\tanh^2 x\sin^2\theta_2\right)\right)}$$

(A7)

The other auxiliary expression is evaluated in a similar way:

$$\left\langle \frac{e^{-2iqN}}{|D|^2} \right\rangle_{N\to\infty} = \frac{1}{2\pi}\int_0^{2\pi} \frac{e^{-i\varphi}d\varphi}{\begin{bmatrix}(\cosh x e^{i\varphi}e^{-i\theta_1}c_+^2 - 2i\sinh x\sin\theta_2 c_+c_- - \cosh x e^{-i\varphi}e^{i\theta_1}c_-^2)\times \\ (\cosh x e^{-i\varphi}e^{i\theta_1}c_+^{*2} + 2i\sinh x\sin\theta_2 c_+^*c_-^* - \cosh x e^{i\varphi}e^{-i\theta_1}c_-^{*2})\end{bmatrix}} =$$

$$\underset{j=\exp(i\varphi)}{=} \frac{-i}{2\pi\cosh^2 x|c_+|^4}\oint_{|j|=1}\frac{dj}{(j-j_1)(j-j_2)(jj_1^*-1)(jj_2^*-1)} =$$

$$= \frac{j_1^* j_2 - j_1 j_2^*}{\cosh^2 x\left(\left|\frac{c_-}{c_+}\right|^2 - 1\right)(j_1-j_2)\left(|c_+|^4+|c_-|^4+2|c_+|^2|c_-|^2\left(1-2\tanh^2 x\sin^2\theta_2\right)\right)} = \tag{A8}$$

$$= \frac{-2ic_+c_-^*\exp(i\theta_1)\tanh x\sin\theta_2}{\cosh^2 x\left(|c_+|^2-|c_-|^2\right)\left(|c_+|^4+|c_-|^4+2|c_+|^2|c_-|^2\left(1-2\tanh^2 x\sin^2\theta_2\right)\right)}$$

**APPENDIX B: EXACT VALUES OF INTEGRALS IN EQ. (25)**

Let us consider the following part from (25) containing polynomial in the denominator,

$$I = \frac{1}{\left(1+4\gamma^2 z^2\right)^2 + 64(m-1)^2\gamma^2 z^6} \tag{B1}$$

Put $\sigma = z^2$

$$I = \frac{1}{64(m-1)^2\gamma^2\left(\sigma^3 + \frac{\gamma^2}{4(m-1)^2}\sigma^2 + \frac{\gamma}{8(m-1)^2}\sigma + \frac{1}{64(m-1)^2\gamma^2}\right)} \tag{B2}$$

If $\sigma_0, \sigma_1, \sigma_2$ are the roots of the polynomial in the denominator of (B2), then,

$$I = \frac{1}{64(m-1)^2\gamma^2(\sigma-\sigma_0)(\sigma-\sigma_1)(\sigma-\sigma_2)} \tag{B3}$$

Let us write the depressed cubic equation by introducing the variable $r$, $\sigma = r - \frac{\gamma^2}{12(m-1)^2}$

(B4)

$$I=\frac{1}{64(m-1)^2\gamma^2\left(\begin{array}{l} r^3+r\left(\frac{1}{8(m-1)^2}-3\left(\frac{\gamma^2}{12(m-1)^2}\right)^2\right)+ \\ \left(2\left(\frac{\gamma^2}{12(m-1)^2}\right)^3-\frac{\gamma^2}{96(m-1)^4}+\frac{1}{64(m-1)^2\gamma^2}\right)\end{array}\right)} \quad \text{(B5)}$$

If we define,

$$\bar{p}=\frac{1}{8(m-1)^2}-3\left(\frac{\gamma^2}{12(m-1)^2}\right)^2;\bar{q}=2\left(\frac{\gamma^2}{12(m-1)^2}\right)^3-\frac{\gamma^2}{96(m-1)^4}+\frac{1}{64(m-1)^2\gamma^2}, \quad \text{(B6)}$$

then one root of the cubic equation can be written as trigonometric and hyperbolic solutions.

$$\sigma_0=\begin{cases} -\frac{\gamma^2}{12(m-1)^2}-2\sqrt{\frac{\bar{p}}{3}}\sinh\left[\frac{1}{3}\operatorname{arcsin}h\left[\frac{3\bar{q}}{2\bar{p}}\sqrt{\frac{3}{\bar{p}}}\right]\right], \text{ if } 4\bar{p}^3+27\bar{q}^2>0 \text{ and } \bar{p}>0 \\ -\frac{\gamma^2}{12(m-1)^2}-2\frac{|\bar{q}|}{\bar{q}}\sqrt{-\frac{\bar{p}}{3}}\cosh\left[\frac{1}{3}\operatorname{arccos}h\left[\frac{-3|\bar{q}|}{2\bar{p}}\sqrt{\frac{-3}{\bar{p}}}\right]\right], \text{ if } 4\bar{p}^3+27\bar{q}^2>0 \text{ and } \bar{p}<0 \\ -\frac{\gamma^2}{12(m-1)^2}+2\sqrt{-\frac{\bar{p}}{3}}\cos\left[\frac{1}{3}\arccos\left[\frac{3\bar{q}}{2\bar{p}}\sqrt{\frac{-3}{\bar{p}}}\right]\right], \text{ if } 4\bar{p}^3+27\bar{q}^2<0 \end{cases} \quad \text{(B7)}$$

Then the remaining roots can be found as follows,

$$I=\frac{1}{64(m-1)^2\gamma^2(\sigma-\sigma_0)(\sigma^2+b\sigma+c)} \quad \text{(B8)}$$

$$b=\frac{\gamma^2}{4(m-1)^2}+\sigma_0;c=\frac{-1}{64(m-1)^2\gamma^2\sigma_0};\sigma_1=\frac{\left(-b+\sqrt{b^2-4c}\right)}{2};\quad \sigma_2=\frac{\left(-b-\sqrt{b^2-4c}\right)}{2}$$

Now, it is easy to do the partial fraction following the exact integration.

$$I=\frac{1}{64(m-1)^2\gamma^2}\left(\begin{array}{l}\frac{1}{(\sigma_1-\sigma_2)(\sigma_1-\sigma_0)(\sigma-\sigma_1)}+ \\ \frac{1}{(\sigma_1-\sigma_2)(\sigma_0-\sigma_2)(\sigma-\sigma_2)}+ \\ \frac{1}{(\sigma_0-\sigma_1)(\sigma_0-\sigma_2)(\sigma-\sigma_0)}\end{array}\right) \quad \text{(B9)}$$

Putting (B9) back to (25) yields

$$J=\frac{8\gamma T}{\pi}\frac{1}{64\left(m-1\right)^{2}\gamma^{2}}\int_{0}^{1}\left(\begin{array}{l}\frac{z^{2}\sqrt{1-z^{2}}\left(1+4\gamma^{2}z^{2}\right)}{\left(\sigma_{1}-\sigma_{2}\right)\left(\sigma_{1}-\sigma_{0}\right)\left(z^{2}-\sigma_{1}\right)}+\\ \frac{z^{2}\sqrt{1-z^{2}}\left(1+4\gamma^{2}z^{2}\right)}{\left(\sigma_{1}-\sigma_{2}\right)\left(\sigma_{0}-\sigma_{2}\right)\left(z^{2}-\sigma_{2}\right)}+\\ \frac{z^{2}\sqrt{1-z^{2}}\left(1+4\gamma^{2}z^{2}\right)}{\left(\sigma_{0}-\sigma_{1}\right)\left(\sigma_{0}-\sigma_{2}\right)\left(z^{2}-\sigma_{0}\right)}\end{array}\right)dz \tag{B10}$$

$$J=\frac{T}{32\left(m-1\right)^{2}\gamma}\left(\begin{array}{l}\frac{1}{\left(\sigma_{1}-\sigma_{2}\right)\left(\sigma_{1}-\sigma_{0}\right)}\left(1+2\sqrt{\sigma_{1}-1}\sqrt{\sigma_{1}}-2\sigma_{1}\right)+\\ \frac{1}{\left(\sigma_{1}-\sigma_{2}\right)\left(\sigma_{0}-\sigma_{2}\right)}\left(1+2\sqrt{\sigma_{2}-1}\sqrt{\sigma_{2}}-2\sigma_{2}\right)+\\ \frac{1}{\left(\sigma_{0}-\sigma_{1}\right)\left(\sigma_{0}-\sigma_{2}\right)}\left(1+2\sqrt{\sigma_{0}-1}\sqrt{\sigma_{0}}-2\sigma_{0}\right)\end{array}\right)+$$
$$\frac{T\gamma}{32\left(m-1\right)^{2}}\left(\begin{array}{l}\frac{1}{\left(\sigma_{1}-\sigma_{2}\right)\left(\sigma_{1}-\sigma_{0}\right)}\left(1+4\sigma_{1}+8\sqrt{\sigma_{1}-1}\sigma_{1}^{3/2}-8\sigma_{1}^{2}\right)+\\ \frac{1}{\left(\sigma_{1}-\sigma_{2}\right)\left(\sigma_{0}-\sigma_{2}\right)}\left(1+4\sigma_{2}+8\sqrt{\sigma_{2}-1}\sigma_{2}^{3/2}-8\sigma_{2}^{2}\right)+\\ \frac{1}{\left(\sigma_{0}-\sigma_{1}\right)\left(\sigma_{0}-\sigma_{2}\right)}\left(1+4\sigma_{0}+8\sqrt{\sigma_{0}-1}\sigma_{0}^{3/2}-8\sigma_{0}^{2}\right)\end{array}\right) \tag{B11}$$

Similarly we can calculate $T_1$ and $T_2$

$$T_{1}=\frac{T}{2}+\frac{T}{16}\left(\begin{array}{l}\frac{1}{\left(\sigma_{1}-\sigma_{2}\right)\left(\sigma_{1}-\sigma_{0}\right)}\left(3+4\sigma_{1}+8\sigma_{1}^{2}-\frac{8\sigma_{1}^{5/2}}{\sqrt{\sigma_{1}-1}}\right)+\\ \frac{1}{\left(\sigma_{1}-\sigma_{2}\right)\left(\sigma_{0}-\sigma_{2}\right)}\left(3+4\sigma_{2}+8\sigma_{2}^{2}-\frac{8\sigma_{2}^{5/2}}{\sqrt{\sigma_{2}-1}}\right)+\\ \frac{1}{\left(\sigma_{0}-\sigma_{1}\right)\left(\sigma_{0}-\sigma_{2}\right)}\left(3+4\sigma_{0}+8\sigma_{0}^{2}-\frac{8\sigma_{0}^{5/2}}{\sqrt{\sigma_{0}-1}}\right)\end{array}\right) \tag{B12}$$

$$T_2 = \frac{T}{2} - \frac{T}{16}\left( \begin{array}{l} \frac{1}{(\sigma_1 - \sigma_2)(\sigma_1 - \sigma_0)}\left( 3 + 4\sigma_1 + 8\sigma_1^2 - \frac{8\sigma_1^{5/2}}{\sqrt{\sigma_1 - 1}} \right) + \\ \frac{1}{(\sigma_1 - \sigma_2)(\sigma_0 - \sigma_2)}\left( 3 + 4\sigma_2 + 8\sigma_2^2 - \frac{8\sigma_2^{5/2}}{\sqrt{\sigma_2 - 1}} \right) + \\ \frac{1}{(\sigma_0 - \sigma_1)(\sigma_0 - \sigma_2)}\left( 3 + 4\sigma_0 + 8\sigma_0^2 - \frac{8\sigma_0^{5/2}}{\sqrt{\sigma_0 - 1}} \right) \end{array} \right) \quad \text{(B13)}$$

(B11)-(B13) give the exact solutions to evaluate Kapitza resistance given in (26).